        \documentstyle[11pt,epsf]{article}
\def\lsim{\,\lower2truept\hbox{${< \atop\hbox{\raise4truept\hbox{$\sim$}}}$}\,}
\def\gsim{\,\lower2truept\hbox{${> \atop\hbox{\raise4truept\hbox{$\sim$}}}$}\,}

\def\deg{\ifmmode^\circ \else$^\circ $\fi}    
\def\arcs{\ifmmode {'' }\else $'' $\fi}     
\def\arcm{\ifmmode {' }\else $' $\fi}     
\def\buildrel#1\over#2{\mathrel{\mathop{\null#2}\limits^{#1}}}
\def\mper{\ifmmode \buildrel m\over . \else $\buildrel m\over .$\fi}
\def\hper{\ifmmode \rlap.^{h}\else $\rlap{.}^h$\fi}
\def\sper{\ifmmode \rlap.^{s}\else $\rlap{.}^s$\fi}
\def\arcsper{\ifmmode \rlap.{' }\else $\rlap{.}' $\fi}
\def\arcmper{\ifmmode \rlap.{'' }\else $\rlap{.}'' $\fi}

\def\mincir{\ \raise -2.truept\hbox{\rlap{\hbox{$\sim$}}\raise5.truept	
\hbox{$<$}\ }}								%
\def\magcir{\ \raise -2.truept\hbox{\rlap{\hbox{$\sim$}}\raise5.truept	%
\hbox{$>$}\ }}								%
\begin{document}
\null

\bigskip
\bigskip
\bigskip
\bigskip
\bigskip
\bigskip
\bigskip
\bigskip

\hspace{2cm} 
{\small  Internal Report ITeSRE/CNR 270/2000}

\vspace{0.5cm} 
\hspace{2cm} 
{\small  March 2000}

\bigskip
\bigskip

\hspace{2cm} {\bf CONSTRUCTION OF A DATABASE}

\hspace{2cm} {\bf OF CMB SPECTRUM OBSERVATIONS}

\medskip

\hspace{2cm} 
{\sc R.~Salvaterra and C.~Burigana}

\vfill

\hspace{2cm} 
{\it Istituto TeSRE/CNR, via P.~Gobetti 101, I-40129 Bologna, Italy}

\bigskip
\bigskip
\bigskip
\bigskip

\bigskip
\bigskip
\bigskip
\bigskip

\newpage

\begin{center}

{\small  Internal Report ITeSRE/CNR 270/2000}

\vspace{0.5cm} 
\hspace{2cm} 
{\small  March 2000}

\bigskip
\bigskip
\bigskip

\large

\hspace{2cm} {\bf CONSTRUCTION OF A DATABASE}

\hspace{2cm} {\bf OF CMB SPECTRUM OBSERVATIONS}

\bigskip
\bigskip

\normalsize

\bigskip

\hspace{2cm} 
{\sc R.~Salvaterra and C.~Burigana}

\bigskip

{\it Istituto TeSRE/CNR, via P.~Gobetti 101, I-40129 Bologna, Italy}

\bigskip
\bigskip
\bigskip
\bigskip
\bigskip
\bigskip

\end{center}

\begin{quotation}

SUMMARY -- We present the complete set of the measures of the CMB absolute 
temperature on the basis on the works published in the 
literature, recognize the main causes of error with the aim of estimating separately 
the magnitude of the systematic and statistical errors, by focussing 
in particular on the most recent observations.
The main purpose of this work is to create a complete 
and reasoned database of CMB absolute temperatures.
The simple database format permits a reading of the data through any text editor or 
through programs for data handling.
This is the first step for a statistical analysis of CMB spectrum data
and their comparison with the theoretical predictions for the distorted spectra
in order to derive constraints on physical processes at very high redshifts.
\end{quotation}

\newpage
 
\section{Introduction}

Since the first observation of the Cosmic Microwave Background (CMB) by 
Penzias \& Wilson (1965), many observations have 
been made to measure the CMB absolute temperature both at high and low frequencies. 
The most precise values are given by the FIRAS instrument aboard the COBE 
satellite (Mather et al. 1999). 
 
In this work we present the complete data of the measurements of the CMB absolute 
temperature. We linger in particular on the most recent observations, and thus 
obtaining a critical database of the measures. We analyse the papers in the 
literature to recognize the main causes of error and to evaluate separately 
the magnitude of the systematical and statistical errors. 
As well known, the CMB spectrum provides significant informations on physical
processes in the universe at very high redshifts (e.g. Danese \& Burigana 1993 
and references therein). 
The construction of a complete and manageable database 
is a first step for a versatile statistical analysis of CMB spectrum data
and their comparison with the theoretical predictions for the distorted spectra.

\medskip
 
For the following discussion, the whole set of the CMB spectrum observations 
has been divided in five sections corresponding to different 
frequencies ranges. Separately, we analyse the measures of the CMB temperature 
obtained from the molecular observations (section~\ref{sez:CN}) and those by FIRAS 
and COBRA (section~\ref{sez:satelliti}). This division in frequency ranges 
is mainly related to the different problems which dominate the determination 
of the CMB 
temperature at different frequencies:

\begin{enumerate} 
 
\item $0.408<\nu<1.0$ GHz (section \ref{sez:0.4-1}), where the uncertainties due to the observation 
site choice and to the determination of the Galactic contribution dominate; 
\item $1.4<\nu<2.0$ GHz (section \ref{sez:1-2}), where the uncertainty on the Galactic temperature dominates; 
\item $2.3<\nu<9.4$ GHz (section \ref{sez:2-9}), where, whilst the Galactic contribution decreases, 
the uncertainty on the atmospheric temperature becomes important in the measure of $T_{CMB}$; 
\item $10<\nu<37$ GHz (section \ref{sez:10-37}), where experiments both from ground and from balloons 
have been made. The last ones allow, observing at an altitude of about 25 km, to reduce the problem 
of the atmosphere contaminating the measuraments from the ground; 
\item above 50 (section \ref{sez:50-}). In this section only the values of the CMB temperature 
from ground and balloon are reported, without further analyses,
since in this range the very accurate results from FIRAS are available. 
\end{enumerate}

\section{General Arguments}

We discuss here some arguments in common to all frequency ranges considered in the following.
In the first subsection we briefly refer to the theorical elements of the 
blackbody spectrum and define the concepts of thermodynamic, 
brightness and antenna temperature. In the second subsection 
we analyse the effect of foreground contamination to the measurements of 
the CMB absolute temperature. This contamination is present at any frequency and is therefore 
preferable to analyse it separately. In the individual sections the impact 
of this contribution is then compared with that introduced by the instrumental noise.

\subsection{The Blackbody Spectrum}

The CMB is expected to be exibit approximately a thermal blackbody (BB) spectrum .
We review here some basic properties of BB spectrum useful for following discussion.

The brightness of a BB spectrum is given by 

$$B_{\nu} = B_{\nu,BB}=\frac{2h\nu^3}{c^2}\eta_{BB},$$

\noindent
where $$\eta_{BB}\equiv\frac{1}{{\rm exp}(h\nu/kT_{th})-1} \, ;$$

here $\nu$ is the observation frequency, $T_{th}$ the true thermodynamic temperature 
and $c$, $k$ and $h$ the speed of light, the Boltzmann and the Planck constants respectively;
$\eta$ is the photon distribution function. 
The photon energy density, $\varepsilon$, and the photon number density, $n$, 
are given by
\begin{eqnarray}
\varepsilon=\frac{8\pi h}{c^3} \int_{0}^{\infty}\eta(\nu)\nu^3\,d\nu \, ; \nonumber
\end{eqnarray}
\begin{eqnarray}
n = \frac{8\pi}{c^3} \int_{0}^{\infty}\eta(\nu)\nu^2\,d\nu \, . \nonumber
\end{eqnarray}

For a BB spectrum $\eta=\eta_{BB}$ and the thermodynamic temperature, $T_{th}$, 
completely determines 
the energy density, $\varepsilon$, and number density, $n$, of the photons.
For a BB spectrum they are given by 

$$\varepsilon_{BB}=aT_{th}^4=\frac{8\pi^5k^4}{15 c^3h^3}T_{th}^4$$

$$n_{BB}=\frac{30\zeta(3)a}{\pi^4k}T_{th}^3 \, ,$$

\noindent
where $\zeta(3)$ is the Riemann Zeta function, $\zeta(3) \simeq 1.20$.

\smallskip
For a generic distribution function $\eta(\nu)$ it is usual to define 
the {\it thermodynamic temperature} 
$T_{th} (\nu)$ at a given frequency $\nu$ as the temperature of a BB with the same 
value of $\eta$ a that frequency $\nu$:
\begin{equation}
\eta (\nu) = [{\rm exp}(h\nu/kT_{th}(\nu))-1]^{-1} \, ,
\end{equation}
or equivalently: 
\begin{equation}
T_{th} (\nu) = \frac{h\nu}{k}\frac{1}{{\rm ln}\left(\frac{1}{\eta (\nu) }+1\right)} \, .
\end{equation}
If $\eta=\eta_{BB}$ then $T_{th}(\nu)$ is exactly 
the thermodynamic temperature $T_{th}$, constant at any frequency.
$T_{th}(\nu)$ is often called also {\it brightness temperature}, $T_{br}(\nu)$.

\bigskip

At low frequencies it is possible to expand $\eta_{BB}$ in powers of 
$h \nu / kT $, so obtaining the approximation
holding in the so-called Rayleigh-Jeans (RJ) region

$$\eta_{BB,RJ}\simeq\frac{kT}{h\nu}\;\; if \;\;\nu\ll\frac{kT}{h} \, .$$

For a generic distribution function $\eta(\nu)$ it is defined also 
the {\it antenna temperature} $T_a (\nu)$ as the temperature at which a BB in the RJ
approximation would  have the same $\eta$ at that frequency $\nu$:

\begin{equation}
\eta (\nu) = \eta_{BB,RJ} (\nu) = \left(\frac{h\nu}{kT_a (\nu)}\right) \, ,
\end{equation}
or equivalently: 
\begin{equation}
T_a (\nu) = \frac{h\nu}{k}\eta (\nu) \, .
\end{equation}

For a BB, in the limit 
$h\nu/kT\ll1$ (i.e. $\nu\lsim$~30~GHz or $\lambda \gsim$~1~cm)
$T_a (\nu) $ is equivalent to $T_{th} (\nu)$. In general 
$T_a (\nu)$ is related to $T_{th} (\nu)$ by 
\begin{equation}\label{eqn:Tth}
T_a(\nu)=T_{th}(\nu) { h\nu/kT_{th}(\nu) \over {\rm exp} (h\nu/kT_{th}(\nu)) -1 }\, .
\end{equation}
$T_a (\nu)$ is frequently called also brightness temperature. 
In this work we will use always the term brightness temperature as equivalent
to the thermodynamic temperature.

\bigskip

\subsection{Foregrounds}\label{sez:foreground}

A relevant issue in the final evaluation of the CMB temperature 
is represented by the necessity of accurately subtracting from the sky temperature 
the contributes from the Galaxy and from unresolved
extragalactic sources [see equation (\ref{eqn:Tcmb})]. In 
particular, the Galactic emission is the main error source at low 
frequencies (see Table \ref{tab:1-2r}), although the level its contribution strongly depends 
on the considered frequency range. Thus, we 
describe separately the valutation and the impact of the Galactic contribution
on the determination of the CMB temperature together with the uncertainty 
associated to its determination.

\medskip

At frequencies below $\sim30$ GHz, the Galactic emission is due principally
to the synchrotron emission and to the thermal bremsstrahlung (free-free).
We refer amply in this section to the treatment of Partridge (1995).

The synchrotron emission is due to the interaction between the Cosmic Rays electrons
and the Galactic magnetic field. Both the energetic electrons and the magnetic
field are largely confined to the disk of our spiral Galaxy, so the synchrotron
radiation is also confined to a band around the sky aligned with the plane of
the Galaxy. The radio frequency spectrum of synchrotron radiation depends on 
the energy spectrum of the relativistic electrons which produce it. If the
energy spectrum is a power law (see  Rybicki \& Lightman 1979, Chapter 6)

$$S_{syn}(\nu)\propto\nu^{(1-\alpha_{syn})/2}$$

\noindent
or

\begin{equation}\label{eqn:syn}
T_{a,syn}(\lambda)\propto \lambda^{(\alpha_{syn} +3)/2}.
\end{equation}

\noindent
In our Galaxy  $\alpha_{syn}$ is close to $2.6$, so 
we have essentially $T_{a,syn}(\nu)\propto\nu^{-2.8}$. 
We analyse below the spectral index determination.

The second mechanism responsible for the radio emission of the Galaxy is thermal
brems\-strahlung. In the regions where the ionized matter is optically thin to 
microwave radiation, the observed antenna temperature produced by plasma at
physical temperature  $T_{pl}$ (of about $10^4$ K for the ionized regions in our
Galaxy) is

$$T_{a,pl}=T_{pl}(1-e^{-\tau})\simeq T_{pl}\tau,$$

\noindent
for small optical depth $\tau$.

To a good approximation, the emission coefficient of ionized matter in the 
radio regime is proportional to  $\lambda^2$. Thus both  $\tau$ and $T_{a,pl}$
are proportional to $\lambda^2$. A more exact treatment shows that the exponent
of the power law dependence lies close to 2.1 in the microwave region 
Partridge 1995).

Synchrotron and bremsstrahlung processes dominate the Galactic emission for all
wavelengths above about 3 mm. At shorter wavelengths, and especially in the
submillimeter region, a new source of emission takes over, namely thermal
emission from warm dust in the Galaxy. The antenna temperature of the warm dust
has a strong wavelength dependence, approximately given by the law

$$T_{a,hot}\propto\lambda^{\gamma},$$

\noindent
with $\gamma$ in the range $-1$ to $-2$. Thus dust emission influences spectral
observations of the CMB only at the shortest wavelength, below 1 mm. 

\begin{figure}
\includegraphics{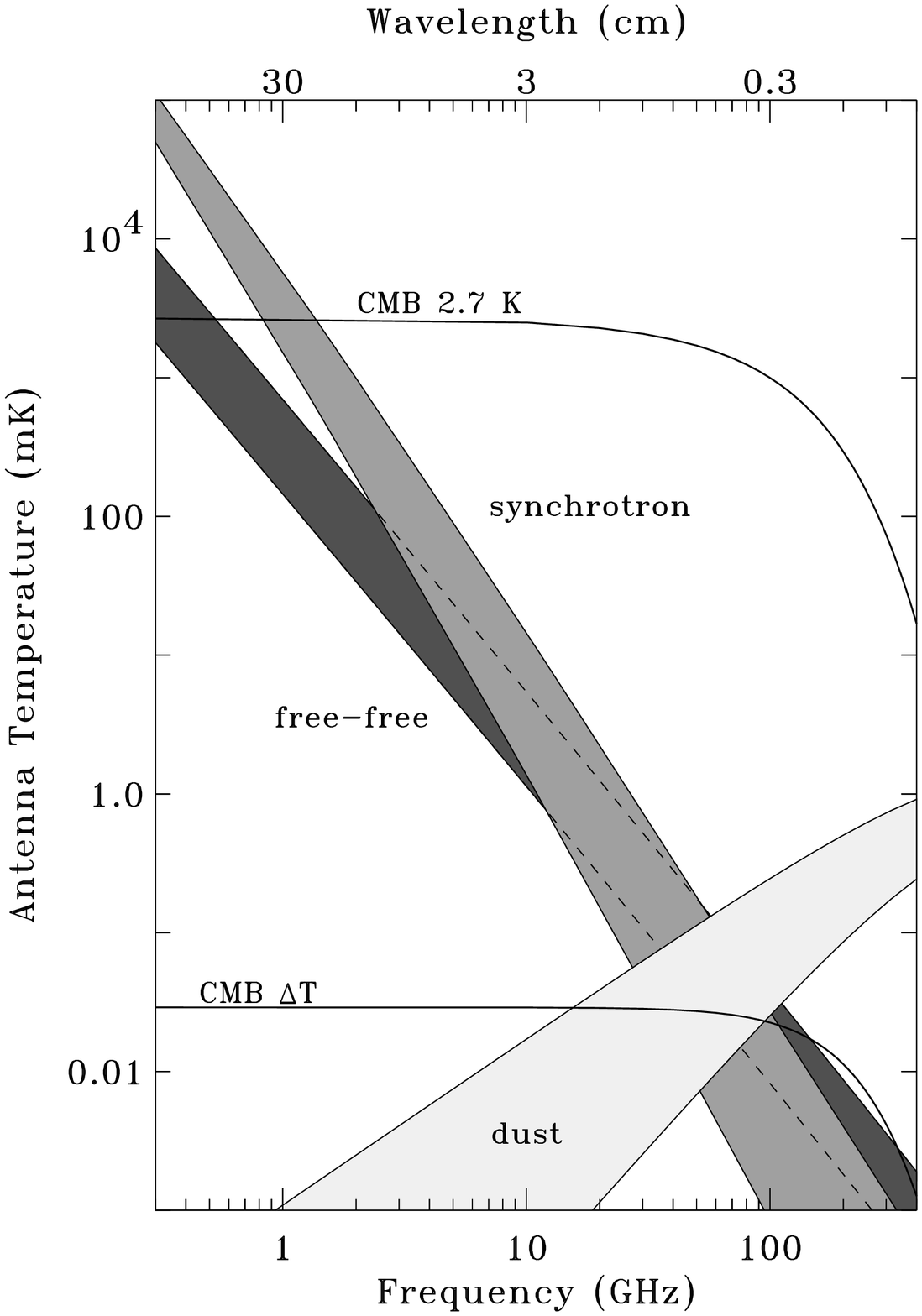}
\caption{Galactic emission components and CMB spectra for moderate angular 
resolution ($7\deg$ \, HPBW) and galactic latitude $|b| <$  $20\deg$. The 
shaded regions indicate the range of synchrotron, free-free and dust emission. 
Solid lines indicate the mean CMB spectrum and rms amplitude of anisotropy
(from Platania et al. 1998).}
\label{contribution}  
\end{figure}

In Figure
\ref{contribution} (Platania et al. 1998) the typical spectra of the 
various sources of the Galactic emission are shown. We note that at 
$6>\lambda>0.07$ cm the Galactic emission is $\leq1$\% of the CMB temperature.
Thus, we expect to have in this region the more accurate measures of the CMB temperature.

\medskip

The Galactic signal is estimated using a 408 MHz skymap (Haslam et al. 1982)
and a compilation of  $H_{II}$ sources at 2.5 GHz. We report in the next
discussion the analysis of Bensadoun (Bensadoun et al. 1993).

The 408 MHz map is first corrected for the CMB signal of $2.7\pm1$ K (which 
allows for the possibility of up to a 1 K spectral distortion at 408 MHz),
for the $H_{II}$ and for the extragalactic sources at 408 MHz [see equation 
(\ref{eqn:Tex})].
The adjusted 408 MHz skymap and the $H_{II}$ map are convolved with the
measured antenna gain pattern to produce a profile at the declination of each
observation. These profiles are then scaled to the frequency of the observation
using spectral indicies of $2.75\pm0.15$ for the 408 MHZ data and $2.10\pm0.05$
for the $H_{II}$ data.

The accuracy of the Galactic model at the observation frequency 
primarily depends on the accuracy of the 408 MHz map and the accuracy of the spectral
index used to scale the map. The Haslam's map is a compilation of four 
different surveys and has overall errors of $\pm3$ K in the zero level and
$\pm10$\% in the gain. At very low frequencies [i.e. 600 MHz (Sironi et al. 1990)],
the uncertainty on the Galactic temperature is dominated from the error on the
zero level of the 408 MHz map. Above 1 GHz, instead, the largest error in the
Galactic signal arises from the uncertainty in the spectral index.
A first approximation of the spectral index comes from the the 408 MHz map 
(Haslam et al. 1982) and
the map at 1420 MHz by Reich \& Reich (1986), after both have been corrected for the
CMB signal. The error on $\alpha_{syn}$ due to the gain and zero level 
uncertainties on the maps at frequencies $\nu_1$ and $\nu_2$ is

$$\delta \alpha_{syn} =\frac{1}{{\rm ln}(\nu_1/\nu_2)}\sqrt{\left(\frac{\delta T_1}{T_1}\right)^2+\left(\frac{\delta T_2}{T_2}\right)^2},$$

\noindent
where $\delta T/T$ is the relative error of the maps.
The $\pm0.5$ K zero level error on the 1420 MHz map dominates the uncertainty 
in $\alpha_{syn}$. An improvement of this evaluation could come by a better
determination of the zero level of the 1420 map (Lawson et al. 1987). 
Otherwise, one can also estimate the spectral index by comparing differences
in the Galactic signal at 408 MHz and at higher frequencies.

Another method for the spectral index valutation consists in producing a 
temperature-temperature plots. Such plots display as ordinate and an abcissa
antenna temperatures measured at two different frequencies on the same region
of the sky. The slope of the best fit to the data, $m$, gives the synchrotron 
index (Platania et al. 1998):

$$\alpha_{syn}=\frac{log(m)}{log(\nu_1/\nu_2)},$$

\noindent
where $\nu_1$ and $\nu_2$ are the frequencies of the two sets of data.

As best valutation of the spectral index, we report here the average from the
analysis of Platania et al. (1998)

$$\alpha _{syn}=2.76\pm0.11$$

Many works discussed here use, however, previous and no so accurate valutations. 

\medskip

In determining the value of the CMB temperature, one must also take into 
account the contribution of the unresolved extragalactic sources (Sironi et al 1990).
This can be described by a power law (Toffolatti \& De Zotti 1988)

\begin{equation}\label{eqn:Tex}
T_{ex}=(23\pm3)\left[\frac{\nu(MHz)}{178}\right]^{-2.75\pm0.05}.
\end{equation}

\bigskip

\section{Range 0.408-1.0 GHz}\label{sez:0.4-1}

The low frequencies measures are due in large part to dated experiments
(see Table \ref{tab:0.4-1}). Also the recent observation (Sironi et al. 1990; Sironi et
al. 1991) give large error bars (Table \ref{tab:0.4-1}), which show no substantial improvement
to previous measures.

\begin{table}[htbp]
\begin{center}
\begin{tabular}{llcl}
\hline
\hline
\multicolumn{1}{c}{$\nu$}  &  \multicolumn{1}{c}{$\lambda$}& $T^{th}_{CMB}$ & Reference \\
\multicolumn{1}{c}{(GHz)}  &  \multicolumn{1}{c}{(cm)} & (K) & \\
\hline
0.408 &  73.5 &  $3.7\pm1.2$ & Howell \& Shakeshaft 1967, Natura, 216, 753 \\
0.6 &  50 & $3.0\pm1.2$ & Sironi et al. 1990, Ap.J., 357, 301 \\
0.610 &  49.1 & $3.7\pm1.2$ & Howell \& Shakeshaft 1967, Natura, 216, 7 \\
0.635 &  47.2 & $3.0\pm0.5$ & Stankevich et al. 1970, Australian J. Phys, 23, 529 \\
0.820 & 36.6 &  $2.7\pm1.6$ & Sironi et al. 1991, Ap.J., 378, 550 \\
1 & 30	& $2.5\pm0.3$ & Pelyushenko \&  Stankevich 1969, Sov. Astron., 13, 223 \\
\hline
\end{tabular}
\end{center}
\caption{Values of the CMB temperature measured at  $\nu\leq 1$ GHz}
\label{tab:0.4-1}
\end{table}

\medskip

An analysis of the recent experiments, made in the last few years, shows that
the large error bars are given from different causes (Table \ref{tab:sironi}) and that
these are due to the different choice of the observation site. We note
however that the uncertainty on the antenna temperature, $T_a$, is greater
than the ones at higher frequencies.

\begin{table}[htbp]
\begin{center}
\begin{tabular}{lcc}
\hline
\hline
 & \multicolumn{2}{c}{$\nu$ (GHz)}\\
\cline{2-3}
Temperature (K) & 0.600 & 0.820 \\
\hline
Antenna ($T_a$) & $16.46\pm0.41$ & $6.7\pm1.5$ \\
Ground ($T_{a,gr}$) & $5.35\pm0.70$ & $0.03\pm0.05$ \\
Atmosphere ($T_{a,atm}$) & $1.17\pm0.30$ & $0.90\pm0.35$ \\
Sun ($T_{a,sun}$) & ... & $0.08\pm0.08$ \\
Galaxy ($T_{a,gal}$) & $6.20\pm0.86$ & $2.67\pm0.33$ \\
Extragalactic & & \\
Sources ($T_{a,ex}$) & $0.81\pm0.11$ & $0.34\pm0.07$ \\
$T^{th}_{CMB}$ & $2.94\pm1.22$ & $2.7\pm1.6$ \\
Site & Alpe Gera & South Pole \\
Reference & Sironi et al. 1990 & Sironi et al. 1991 \\
\hline
\end{tabular}
\end{center}
\caption{Values and errors of the recent experiments}
\label{tab:sironi}
\end{table}

\medskip

For  experiments in Table \ref{tab:sironi} the absolute measurement of the CMB are done by
comparing the output of a radiometer when looking the sky and when looking at a
precisely known calibration blackbody maintained at cryogenic temperature, and
by subtracting-out any other non cosmological contribution (Bonelli et al. 
1995). 
 
The antenna temperature of the sky $T_a$ is calculated by adding the effective 
temperature $T_{eff}$ of the blackbody calibrator and the measured difference 
$\Delta T = G(V_a - V_{cs})$, where $V_a$ and $V_{cs}$ are the signals produced respectively by the 
sky and by the calibrator and $G$ is the system gain (Bonelli et al. 1995). 
More precisely (Sironi et al. 1991):

\begin{equation}\label{eqn:Ta}
T_a = \left[\frac{T_{eff}+\Delta T - rT_r}{1-r}-T_0(1-e^{-\tau})\right]/ e^{-\tau},
\end{equation}
 
\noindent 
where $T_r$ is the temperature of the noise radiated by the system, $r$ is 
the power reflection coefficient of the antenna, $e^{-\tau}$ the trasmission 
coefficient of the line (cable plus horn) which brings the signal from the 
antenna to the receiver and $T_0$ is the physical temperature of the system. 
 
The ideal cold source is a blackbody with the same sky temperature and 
the closest approximation of this could be represented by an optically thick 
absorber inside a dewar 
filled with liquid helium which fits the antenna mouth (Sironi et al. 1990). 
At low frequencies, however, that solution cannot be used because of the 
antenna dimensions. It's used instead a coaxial termination, immersed in liquid 
helium and coupled to the receiver input via cable. The effective temperature 
$T_{eff}$ is given then (Sironi et al. 1990): 
 
\begin{equation}\label{eqn:teff}
T_{eff}=T_l+(\langle T \rangle-T_l)y+T_rR_c,
\label{eqn:Teff}
\end{equation}
 
\noindent
where $T_l$ is the boiling temperature of the liquid in the dewar, $\langle T \rangle$ is the 
average temperature of the cable between the termination and the receiver, 
$T_r$ is the noise temperature radiated by the receiver, $y$ is the power 
absorption coefficient of the cable, and $R_c$ is the power reflection 
coefficient of the cable. $T_l$ depends on  the ambient pressure $p$  and is 
well known when $p$ is known. $T_r$, $y$ e $R_c$ can be measured. Calculating 
the equation (\ref{eqn:Teff}) the result is that the effective temperature of the coaxial 
source is higher than that of the liquid helium. 
 
The most convenient configuration, suggested by equation (\ref{eqn:Ta}), is the one 
in which the source is coupled to the receveir through the same antenna 
which looks at the sky, so that $e^{-\tau}=1$ (Sironi et al. 1991). That is not 
possible for the observations at these frequencies because of the antenna 
size. The error on $T_a$ is also dominated by the uncertainty of $e^{-\tau}$. 
$e^{-\tau}$ is the product of $\epsilon_f$, $\epsilon_w$ and $\epsilon_c$. 
There are respectively the flaring section coefficient, the antenna 
waveguide section coefficient and the coaxial components coefficient. At 
the South Pole (Sironi et al. 1991) because of the severe environmental 
conditions it's not possible to remeasure $\epsilon_w$ at the observing site 
before the waveguide section is fixed at the 
antenna throat. The value of $\epsilon_f$ is given by the average of the values measured in the 
laboratory before and after the experiment. There is also an additional uncertainty 
produced by repeated assembling and disassembling of the waveguide section 
of the horn. Also with reference to the South Pole experiment (Sironi et al. 1991)  
an undesired source of uncertainty of statistical type comes from the 
connectors at the end of the cable, placed between the antenna and the 
receiver to overcome the stiffness of the lines produced by the low 
ambient temperature. 
 
Anyway, a simple comparison with observations in higher frequencies shows 
that the errors 
on $\Delta T$ and $T_{eff}$ (and through equation \ref{eqn:Ta} on $T_a$) are increased by the 
impossibility to fit the antenna mouth  with the cold source. Table \ref{tab:0.8v2.5}
shows a comparison between the uncertainties of these terms measured in the 
same observation project. It is evident that the different tecnological 
solution used at 2.5 GHz (Sironi et al. 1991) allows to reduce the error bars 
on $\Delta T$ and $T_{eff}$. 

\begin{table}[htbp]
\begin{center}
\begin{tabular}{lccc}
\hline
\hline
 & \multicolumn{3}{c}{$\nu$ (GHz)}\\
\cline{2-4}
  & 0.600 & 0.820 & 2.5 \\
\hline
$\Delta T$ & $\pm0.28$ & $\pm0.39$ & $\pm0.01$ \\
$T_{eff}$ & $\pm0.30$ & $\pm0.42$ & $\pm0.01$ \\
Site & Alpe Gera &\multicolumn{2}{c}{South Pole} \\
Reference & Sironi et al. 1990 &\multicolumn{2}{c}{Sironi et al. 1991} \\
\hline
\end{tabular}
\end{center}
\caption{Comparation of the uncertainties on $\Delta T$ and $T_{eff}$ expressed in K at 0.60 GHz, 0.82 GHz and 2.5 GHz}
\label{tab:0.8v2.5}
\end{table}

\medskip
 
From the antenna temperature must be subtracted the not cosmological 
contributions. There are local and foreground contributions. 
 
The sky temperature, $T_{a,sky}$, is achieved by the subtraction of the local terms 
from the antenna temperature $T_a$. More precisely (Sironi et al. 1991): 

\begin{equation}\label{sky}
T_{a,sky}=T_a-T_{a,gr}-T_{a,atm}-T_{a,sun},
\end{equation}
 
\noindent
where $T_{a,gr}$, $T_{a,atm}$, $T_{a,sun}$ are the contributions from the ground 
around the antenna, the Earth's atmosphere and the Sun, respectively.

$T_{a,gr}$ depend on the observation site (Table \ref{tab:sironi}). This term is calculated 
by convolving the antenna beam profile with a blackbody radiator at 
ambient temperature having the shape of the ground visible (i.e., not 
hidden by screens) from the horn mouth (Sironi et al. 1991). At the South 
Pole this is not a problem because the ground is flat and the ambient 
temperature is very low. At Alpe Gera (Sironi et al. 1990), on the 
contrary, because of 
the horizon shape, very large screens are necessary to stop the ground 
radiation, but the screen efficiency is questionable because of the 
presence of few incumbent peaks which make diffraction of their thermal 
radiation over the screen edges highly probable but difficult to evaluate 
and/or cut. Thereafter the big error bars of $T_{a,gr}$ in the experiment at 
Alpe Gera obtain an explanation (Table \ref{tab:sironi}). 
 
On the contrary the Sun contribution must be taken into account only in the South Pole 
observations, where is not possible to measure during the night. To reduce 
$T_{a,sun}$, the radiometers are sorrounded on three sides by reflecting screens. The 
resulting value of $T_{a,sun}$ is calculated taking into account the diffraction 
of the radiation at the screen edges and the antenna power pattern. 
 
The contribution of the Earth's atmosphere can be obtained in principle 
by measuring the variation of the antenna temperature when the horn is 
tilted from zenith angle $z_1$ to  $z_2$. At low frequences, however, the 
galactic signal overcomes the atmospheric signal by a factor ranging from 
3 - 30, depending on the observing direction (Sironi et al. 1990). 
As a result, the value one finds by scaling data at other frequencies is 
preferable compared to the directly measured value of $T_{a,atm}$. Below a few GHz, 
the atmosphere is optically thin, and its emission is dominated by the oxygen 
content of the air mass above the antenna. Therefore, at a given frequency 
$T_{a,atm}$ can be scaled from one observing site to another by the ratio of the 
atmospheric pressure at the two sites. To compare data at different 
frequencies, the dependence of $T_{a,atm}$ on $\nu$ is necessary (Sironi et al. 1990). 
Below a few GHz most atmospheric models (i.e., Danese \& Partridge 1989) 
predict a nearly flat spectrum down to about 1 GHz and a decrease somewhere 
below, although is not possible exclude that the spectrum may remain 
flat down to 500 MHz. Assuming a flat spectrum, we have a value for 
$T_{a,atm}$. The error bar has been set large enough to be compatible with the 
models and the observational data at nearby frequencies. 

\bigskip
 
The CMB temperature is obtained by subtracting the foreground contributions 
from the sky temperature, $T_{a,sky}$: 
 
\begin{equation}\label{eqn:Tcmb}
T_{a,CMB}(\nu)=T_{a,sky}(\alpha,\delta,\nu)-T_{a,gal}(\alpha,\delta,\nu)-T_{a,ex}(\nu), 
\end{equation}

\noindent 
where $\alpha$ and $\delta$ are the celestial coordinates (Sironi et al. 1990).
 
Between $\sim$~0.5 and 1 GHz, none of the three components of $T_{a,sky}$ is 
neglegible. In particular, at 600 MHz outside the galactic plane, one 
expects the following proportions (Sironi et al. 1990): 

$$T_{a,gal}:T_{a,CMB}:T_{a,ex}\simeq 7:3:1.$$
  
The methods to determine $T_{a,gal}$ and $T_{a,ex}$ are already described in the subsection (\ref{sez:foreground}). 
We only note (Sironi et al. 1990) that in this range of frequencies 
the uncertainty on $T_{a,gal}$ is dominated from the error on the 408 MHz map 
(Haslam et al. 1982). Were $T_{a,sky}(408)$ known with the accuracy of 0.9K, 
the error bars could be reduced by about a factor of 2. In this case, the 
1 K uncertainty of $T_{a,CMB}(408)$ would dominate (Sironi et al. 1991). 

\bigskip
 
By subtracting all the non cosmological contributes the antenna temperature 
of CMB radiation is derived; its error bars are obtained by adding in 
quadrature the uncertainties on the terms of equations (\ref{eqn:Ta}), (\ref{sky}) and (\ref{eqn:Tcmb}). 
 
Through equation (\ref{eqn:Tth}) we can turn the antenna temperature to the 
thermodynamic temperature $T_{th,CMB}$, although the difference between these two 
is less than 1\% at these low frequencies. 
 
\bigskip

\section{Range 1.4-2.0 GHz}\label{sez:1-2}

Many measures in this range are carried out in recent times. 
One expects that observations in this range are low enough in frequency 
to allow significant distortions, but high enough that the Galactic 
signal is still a factor of 3 weaker than the CMB (Bensadoun et al. 1993). 
Table \ref{tab:1-2} gives the data in this range, while Table \ref{tab:1-2r} shows a comparation 
of the recent measures. It can be noted that the main error source derives from 
the determination of the Galactic temperature. 

\begin{table}[htbp]
\begin{center}
\begin{tabular}{llcl}
\hline
\hline
\multicolumn{1}{c}{$\nu$}  &  \multicolumn{1}{c}{$\lambda$}& $T^{th}_{CMB}$ & Reference \\
\multicolumn{1}{c}{(GHz)}  &  \multicolumn{1}{c}{(cm)} & (K) & \\
\hline
1.4 &  21.3 & $2.11\pm0.38$ & Levin et al. 1988 \\
1.42 & 21.2 & $3.2\pm1.0$ & Penzias \& Wilson, 1967 \\
1.43 & 21 & $2.65^{0.33}_{0.30}$ & Staggs et al. 1996 \\
1.44 & 20.9 & $2.5\pm 0.3$ & Pelyushenko \& Stankevich 1969  \\
1.45 & 20.7 & $2.8\pm0.6$ & Howell \& Shakeshaft 1966 \\
1.47 & 20.4 & $2.27\pm0.19$ & Bensadoun et al. 1993 \\
2 & 15 & $2.5\pm0.3$ & Pelyushenko \& Stankevich 1969 \\
2 & 15 & $2.55\pm0.14$ & Bersanelli et al. 1994 \\
\hline
\end{tabular}
\end{center}
\caption{Values of the CMB temperature at frequencies $1<\nu\leq 2$ GHz}
\label{tab:1-2}
\end{table}

\medskip

\begin{table}[htbp]
\footnotesize
\begin{center}
\begin{tabular}{lcccc}
\hline
\hline
 & \multicolumn{4}{c}{$\nu$ (GHz)}\\
\cline{2-5}
Temperature (K) & 1.410 & 1.43 & 1.47 & 2 \\
\hline
$G(S_a-S_{load})$ & $-0.06\pm0.03$ & ... & $0.39\pm0.02$ & $0.17\pm0.03$ \\
Source ($T_{a,load}$) & $3.78\pm0.31$ & ... & $3.86\pm0.03$ & $3.85\pm0.03$ \\
Instruments ($\Delta T_{a,inst}$) & $0.0\pm0.083$ & $0.07\pm0.02$ & $0.02\pm0.06$ & $0.07\pm0.04$ \\
Joint ($\Delta T_{a,joint}$) & ... & $0.79^{+0.15}_{-0.21}$$^a$ & $0.025\pm0.070$ & ... \\
Atmosphere ($T_{a,atm}$) & $0.83\pm0.10$ & $1.82\pm0.20$ & $1.08\pm0.08$ & $1.07\pm0.07$ \\
Galaxy ($T_{a,gal}$) & $0.80\pm0.16$ & $5.23\pm0.16$$^b$  & $0.82\pm0.19$ & $0.33\pm0.10$ \\
Ground ($T_{a,gr}$) & $0.017\pm0.008$ & $0.040\pm0.025$ & $0.06\pm0.03$ & $0.05\pm0.04$ \\ 
RFI ($T_{a,RFI}$) & ... & ... & $0.0\pm0.01$ & $0.0\pm0.01$ \\
Sun ($T_{a,sun}$) & ... & ... & $<0.01$ & $<0.01$ \\
$T^{th}_{CMB}$ & $2.11\pm0.38$ & $2.65^{+0.33}_{-0.30}$ & $2.26\pm0.20$ & $2.55\pm0.14$ \\
Sito & White Mountain & West Virginia & WM/SP & South Pole \\
Reference & Levin, 1988 & Staggs, 1996 & Bensadoun, 1993 & Bersanelli, 1994 \\
\hline
\end{tabular}
\end{center}
\caption{Values and errors of the recent experiments}
\label{tab:1-2r}
$^a$ The reported value is the emission of the system of the waveguide and the joint (Staggs et al. 1996)

$^b$ In this table is reported the value of $T_a-T_{a,gal}$, as given in the work of Staggs et al. (1996) 
\end{table}

\medskip
 
The experiment consists in comparing the signal from the zenith $S_a$ with one 
coming from a large, liquid-helium cooled, cold-load calibrator $S_{load}$ whose antenna 
temperature, $T_{a,load}$, is precisely known. The antenna temperature of the 
zenith, $T_{a,zenith}$, is 

\begin{equation}\label{eqn:1-2:Ta}
T_{a,zenith}=G(S_a-S_{load})+T_{a,load}-\Delta T_{a,inst}-\Delta T_{a,joint},
\end{equation}
 
\noindent
where $G$ is the gain calibration coefficient for the radiometer, $S_a-S_{load}$ 
is the measured signal difference between the zenith and the cold load, 
$\Delta T_{a,inst}$ is the correction for any changes in the radiometer signal 
associated with its inversion during the measurement, and $\Delta T_{a,joint}$ is the 
differential temperature contribution from the imperfect joint between the 
antenna and the cold load (Bensadoun et al. 1993). 

\bigskip
 
The cold source calibrator is an optical thick absorber in a LHe-cooled dewar 
(Smoot et al. 1983), that works like a 
waveguide fitting the antenna mouth, so that in equation (\ref{eqn:Ta}) $e^{-\tau}$ becomes $\simeq 1$. 
Errors in the measure of $T_{a,load}$ are due to the fact that the absorbing material is 
not infinitely thick and also due to the interface between the radiometer and 
the dewar. As a result, a small fraction of the power emitted by the 
radiometer is reflected back into the antenna, contributing to the antenna 
temperature of the absolute reference load (Levin et al. 1988). In recent 
experiments (Bensadoun et al. 1993) this effect is minimized. 
Particularly in the observation of 1988 (Levin et al. 1988) this term 
contributes in large part to the final error on $T_{a,CMB}$, whereas in 
recent experiments the uncertainties on $T_{a,load}$ are very low (see 
Table \ref{tab:1-2r}). 
 
As stated above, the difference $G(S_a-S_{load})$ is measured 
repeatedly during the experiment and then follows a Gaussian parent distribution. 
The error is based solely on the 
statistical fluctuations (Levin et al. 1988). This is however smaller 
compared to systematic errors set, that contaminates the observations. 
The gain variation does not follow a Gaussian distribution, but in fact 
shows slow drifts (Levin et al. 1988). The horn emission is neglected.  
 
The instrumental offset term in equation (\ref{eqn:1-2:Ta}) takes in consideration the fact 
that the radiometer properties can change when it is inverted to observe the 
cold load (Bensadoun et al. 1993). Changes in the gain $\delta G$, the system 
temperature $\delta T_{sys}$, the physical temperature of any loss front-end components  $\delta T_R$ and 
the reflection  $\delta R$ and the insertion loss $\delta L$ of each component can contribute 
to the global instrumental uncertainty $\Delta T_{inst}$. More precisely (Bensadoun et al. 1993): 

\begin{eqnarray*}
\Delta T_{a,inst} & \simeq &  \delta T_{sys}+\delta T_B R+\delta T_R L- \\
                & & \mbox{}-\frac{\delta G}{G} [T_{sys}+T_{a,load}(1-R-L)+T_B R+T_R L]+ \\
                & & \mbox{}+\delta R(T_B-T_{a,load})+\delta L(T_R-T_{a,load}),
\end{eqnarray*}
 
\noindent
where $T_B$ is the broadcast noise temperature. 
 
To the term $\Delta T_{a,joint}$ in equation (\ref{eqn:1-2:Ta}) contribute the differences in the 
joints between the antenna horn and the interface plates of the cold load 
and of the ground screen. 
$\Delta T_{a,joint}$ is due to the differential emission from within the resistive 
metal to metal joints, the differential transmission of ambient radiation 
through the joints and the differential joint reflection. The first two 
terms carry the larger part of the error on $\Delta T_{a,joint}$ because the test is 
not performed with a LHe temperature absorber and the value comes out from 
the test made with the load absorber immersed in liquid nitrogen. The 
difference in reflection between the antenna-ground screen interface and the 
antenna-load interface is measured using a network analyzer. 

\bigskip
 
The CMB temperature is obtained by subtracting from the antenna temperature  
all non cosmological terms: 

\begin{equation}\label{eqn:1-2:Tcmb}
T_{a,CMB}=T_a-T_{a,for}-T_{a,atm}-T_{a,gr}-T_{a,sun}-T_{a,RFI},
\end{equation}
 
\noindent
where $T_{a,for}$ is the foregrounds contribution, $T_{a,atm}$ is the atmosphere 
contribution, $T_{a,gr}$ is the ground radiation contribution, $T_{a,sun}$ is the Sun 
contribution and $T_{a,RFI}$ is the manmade interference contribution. 
 
The foreground term $T_{a,for}$ comes by adding the Galactic radiation $T_{a,gal}$ and 
the extragalactic source radiation $T_{a,ex}$, formerly discussed in the subsection 
\ref{sez:foreground}. The extragalactic source contribution is neglected here, because it is small 
respect to the Galactic component. We note here, that at these frequencies 
the error on $T_{a,gal}$ is the greatest cause of the uncertainty on $T_{a,CMB}$. The 
Galactic signal is estimated using a 408 MHz skymap (Haslam et al. 1982), 
scaled at the observation frequency (Bensadoun et al. 1993). The largest error 
in the Galactic signal arise from uncertainty in the spectral index $\alpha$, 
valued by the difference in the Galactic signal at 408 MHz 
and at higher frequencies. 
 
$T_{a,gr}$ and $T_{a,sun}$ are described in the subsection above. 
 
$T_{a,RFI}$ is controlled by the continuous monitoring of the man-made 
interference. This is an important factor in selecting the observation site. 
In conclusion, this term introduces only a small change and therefore its 
contribution is negligible (Bensadoun et al. 1993). 
 
It remains only to be analysed the atmospheric contribution. At frequencies 
below 2 GHz the simple extrapolation from values measured at higher 
frequencies is favourable rather than a direct measurement (Ajello et al. 
1995; Levin et al. 1988; Bensadoun et al. 1993). The atmospheric signal 
in this range is due to resonant and nonresonant emission by complex of 
pressure-broadened oxygen lines clustered near $\nu=60$ GHz. The amplitude 
of the $O_2$ emission depends on atmospheric pressure and temperature 
(Bensadoun et al. 1993). The water vapour is instead negligible. Over the 
range $1 < \nu < 10$ GHz in the simple, dry atmosphere model of Gordon (1967), 
the attenuation, $\alpha$, scales approximately as 

$$\alpha = A\gamma \frac{x(1+3x)}{g(1-3x)^2+x(1-x)^2},$$
 
\noindent
where $x=(\nu / \nu_0)^2$, $g=(\gamma / \nu_0)^2$, and $A$ and $\gamma$ are the pressure- and temperature-dependent 
amplitude and line width parameters for oxygen. In the extrapolation the 
effect of the finite beam and the uncertain on the parameter $\gamma$ has been 
taken into account. The simple extrapolation from values measured at nearby 
frequencies agrees well with the empirical atmospheric attenuation model of 
Danese \& Partridge (1989). 
 
The error on $T_{a,atm}(0)$ for this method comes from the accurancy of the measure 
of $p$, $T$ and the humidity $u$ of the observation site, from the fluctuations 
of the atmospheric conditions, from the uncertain on $\alpha$ and from the 
change of $p$, $T$ and $u$ if there are obtained from balloon flights (Ajello 
et al. 1995). 
 
A cross-check of this value comes from a direct measurement of 
the atmospheric temperature. At around 1.5 GHz, the Galactic background is 
no longer so overhanging to the atmospheric signal. 
Two methodes are used to determine $T_{a,atm}(0)$ (Ajello et al. 1995). 
By the {\itshape estinction method} one measures the 
dependence of the apparent temperature of a bright source on $\theta$, the 
zenith angle. Then, assuming an effective temperature, the atmospheric 
temperature at zenith can be obtained. This method produces error bars ranging 
between 0.02 K below 1 GHz and 0.2 K above 1 GHz and has been used at low 
frequencies. By the {\itshape emission method} one assumes a constant sky signal and 
measure $\Delta T(\theta_1,\theta_2)$, the variation of the atmospheric noise when the 
antenna zenith angle goes from $\theta_1$ and $\theta_2$. We have also 

$$T_{a,atm}(0)=\frac{\Delta T(\theta_1,\theta_2)}{\langle f(\theta_1) \rangle - \langle f(\theta_2) \rangle},$$
 
\noindent
where $\Delta T$ and $f(\theta)$ are averaged over the antenna 
beam\footnote{See section \ref{sez:2-9} for a more accurate analysis}. 
At frequencies  $\leq$3.8 GHz  the systematic errors dominate (Bersanelli et al. 1995). 
 
Bersanelli et al. (1994) carried out a set of measures at different 
frequencies. In Table \ref{tab:m-e} are the values obtained and the extrapolations 
at 1.47 and 2 GHz. The measured value and 
that extrapolated at 2 GHz are in good agreement. 

\begin{table}[htbp]
\begin{center}
\begin{tabular}{lccc}
\hline
\hline
Frequency & Measured & Estrapolated & Year\\
(GHz) & (K) & (K) & \\
\hline 
7.5 & $1.174\pm0.064$ & ... & 1989 \\
3.8 & $1.070\pm0.060$ & ... & 1989 \\
2.0 & $0.989\pm0.070$ & $1.020\pm0.070$ & 1991 \\
1.47 & ... & $0.977\pm0.068$ & ... \\
\hline
\end{tabular}
\end{center}  
\caption{Comparation between the measured and the estrapolated atmospherical emission (South Pole; Pencil beam value) from Bersanelli et al. (1994)}
\label{tab:m-e}
\end{table}

\medskip
 
By subtracting all contributions we derive $T_{a,CMB}$. The error bars are obtained 
by adding in quadrature the uncertainties on the terms in equation (\ref{eqn:1-2:Ta}) and (\ref{eqn:1-2:Tcmb}).
Through equation (\ref{eqn:Tth}) we converte $T_{a,CMB}$ to thermodynamic temperature $T^{th}_{CMB}$, 
whose error comes from the propagation of errors.

\bigskip

\section{Range 2.3-9.4 GHz}\label{sez:2-9}

In the observations between 2.3 and 9.4 GHz (see Tables \ref{tab:2-9} and \ref{tab:2-9r}) it is 
clear that the uncertainty on the foreground contribution on the final 
error decreases when the frequency increases; whilst the error on the 
atmospheric temperature dominates and represents the major cause of the 
width of the error bars above 2 GHz. As in other subsections, we analyse 
here the most recent experiments (Table \ref{tab:2-9r}). We note however the value of 
Stokes et al. (1967) that is in good agreement with the FIRAS results (see 
\ref{sez:FIRAS}) and that has a very low margin of error. 
 
\begin{table}[htbp]
\begin{center}
\begin{tabular}{llcl}
\hline
\hline
\multicolumn{1}{c}{$\nu$}  &  \multicolumn{1}{c}{$\lambda$}& $T^{th}_{CMB}$ & Reference \\
\multicolumn{1}{c}{(GHz)}  &  \multicolumn{1}{c}{(cm)} & (K) & \\
\hline
2.3 & 13.1 & $2.66\pm0.7$ & Otoshi \& Stelzreid 1975, IEEE Trans on Inst \& Meas, 24 \\
2.5 & 12 & $2.71\pm0.21$ & Sironi et al 1991, Ap.J., 378, 550 \\
3.8 & 7.9 & $ 2.64\pm0.06$ & De Amici et al 1991, Ap.J., 381, 341 \\
4.08 & 7.35 & $3.5\pm1.0$ & Penzias \& Wilson, 1965, Ap.J., 142, 419 \\
4.75 & 6.3 & $2.70\pm0.07$ & Mandolesi et al. 1986, Ap.J., 310, 561 \\
7.5 & 4.0 & $2.60\pm0.07$ & Kogut et al. 1990, Ap.J., 355, 102 \\
7.5 & 4.0 & $2.64\pm0.06$ & Levin et al. 1992, Ap.J.,396, 3 \\
9.4 & 3.2 & $3.0\pm0.5$ & Roll \& Wilkinson 1966, PRL, 16, 405 \\
9.4 &  3.2 & $2.69^{+ 0.16}_ {-0.21}$ & Stokes et al. 1967, PRL, 19, 1199 \\
\hline
\end{tabular}
\end{center}
\caption{Values of the CMB temperature at $2.3\leq \nu \leq 9.4$ GHz}
\label{tab:2-9}
\end{table}

\medskip

\begin{table}[h]
\tiny
\begin{center}
\begin{tabular}{lcccccc}
\hline
\hline
 & \multicolumn{6}{c}{$\nu$ (GHz)}\\
\cline{2-7}
Temperature (K) & 2.5 & 3.8 & 4.75 & 7.5 & \multicolumn{2}{c}{7.5} \\
 & & & & & 1988 & 1989 \\
\hline
$G(S_a-S_{load})$ & ... & $-0.009\pm0.008$ & $-0.045\pm0.013$ & ... & $-0.146\pm0.012$ & $-0.126\pm0.013$ \\
Source ($T_{a,load}$) & $3.73\pm0.15$ & $3.762\pm0.019$ & $3.682\pm0.010$ & $3.621\pm0.009$ & $3.671\pm0.023$ \\
Atmosphere ($T_{a,atm}$) & $1.155\pm0.300$ & $1.109\pm0.060$ & $0.997\pm0.060$ & $1.083\pm0.055$ & $1.083\pm0.059$ & $1.222\pm0.064$ \\
Galaxy ($T_{a,gal}$) & $0.118\pm0.025$ & $0.055\pm0.015$ & $0.035\pm0.025$ & $0.010\pm0.005$ &  $0.010\pm0.005$ & $0.007\pm0.004$ \\
Ground ($T_{a,gr}$) & $0.030\pm0.050$ & $0.006\pm0.008$ & $0.020\pm0.010$ & $0.013\pm0.010$ & $0.013\pm0.010$ & $0.022\pm0.015$ \\
System ($\Delta T_{sys}$) & ... & $0.034\pm0.034$ & $0.0\pm0.020$ & $0.052\pm0.034$  & $0.052\pm0.034$ & $0.023\pm0.025$ \\
RFI ($T_{a,RFI}$) & ... & ... & ... & $0.0\pm0.005$ & ... & $0.0\pm0.005$ \\
Sun ($T_{a,sun}$) & $0.0\pm0.005$ & ... & ... & ... & ... & ... \\
$T_{a,ex}$ & $0.016\pm0.005$ & ... & ... & ... & ... & ... \\
$T^{th}_{CMB}$ & $2.50\pm0.34$ & $2.64\pm0.06$ & $2.70\pm0.07$ & $2.60\pm0.07$ & \multicolumn{2}{c}{$2.64\pm0.06$} \\
Site & SP & WM/SP & WM & WM & WM & SP \\
Reference & Sironi, 1991 & De Amici, 1991 & Mandolesi, 1986 & Kogut, 1990 & \multicolumn{2}{c}{Levin 1992} \\
\hline
\end{tabular}
\end{center}
\caption{Values and errors of the recent experiments}
\label{tab:2-9r}
\end{table}

\medskip
 
The antenna temperature is obtained by the comparation of the sky signal 
with the signal 
of a cold well known load, made by an optical thick absorber inside a 
dewar cooled with liquid helium, that acts like an overmode waveguide which 
fits the antenna mouth (Sironi et al. 1991). 
 
The antenna temperature of CMB is given by the subtraction of the non 
cosmological contributions: 

\begin{equation}\label{eqn:2-9:Tacmb}
T_{a,CMB}=G(S_a-S_{load})+T_{a,load}-T_{a,gal}-T_{a,atm}-T_{a,gr}-T_{a,RFI}-\delta T_{sys},
\end{equation}
 
\noindent
where the terms are already explained in section \ref{sez:1-2}. The error bars on 
$T_{a,CMB}$ is given by adding in quadrature of the uncertainties of the terms in 
equation (\ref{eqn:2-9:Tacmb}). 
 
Again, we pass to the thermodynamic temperature through the equation (\ref{eqn:Tth}).

\medskip

As far as the non cosmological contributions, we must examine in detail only 
the atmospherical and system contributions, since the others are already explained 
in the previous sections (see subsection \ref{sez:foreground} for $T_{a,gal}$ and section \ref{sez:1-2} for the others). 
 
\medskip

The term $\delta T_{sys}$ refers to any systematic change (such as might be caused 
by gravitational stresses or twisting cables) in the radiometer performance 
when it is inverted to look at the cold-load calibrator. We have (De Amici 
et al. 1991): 

\begin{equation}\label{tsys}
\delta T_{sys}=\delta T_{a,inst}+\frac{\delta G}{G}(T_{a,load}+T_{sys})+\delta R(T_{sys}-T_{a,load})+\delta L(T_{phys}-T_{a,load}),
\end{equation}
 
\noindent
where $\delta G/G$ represents the fractional change in calibration coefficient 
between the upright and the upside down position, $\delta R$ is the change in 
reflection coefficient of the horn and amplifier, $\delta L$ is the change in 
insertion loss (attenuation of the incoming signal) of the radiometer, 
$T_{phys}$ is the physical temperature of the components, $T_{sys}$ is the system 
temperature of the receiver and $\Delta T_{inst}$ is the position-dependent 
change in receiver output. This term is measured (Levin et al. 1982) by 
'flip tests', in which a stable target is fixed to the radiometer, which 
is repeatedly inverted to simulate CMB measurement. It is not possible however to build 
an invertible liquid helium target; therefore the value of $\delta T_{sys}$ at 
this temperature can only be extrapolated from the measurements at higher 
temperatures. Contributions to $\delta T_{sys}$, given from changes in insertion loss 
or from reflection, should be proportional to the difference between the 
target and radiometer temperature. Contributions, given by 
changes in gain, should be proportional to the target temperature plus 
radiometer system temperature. Electrical effects should be independent 
of target temperature. The sum, $\delta T_{sys}$, therefore, should be linear 
in the target temperature. The value and the error of $\delta T_{sys}$ at 2.7 K 
is calculated minimizing the $\chi^2$ of two-dimensional linear fit. 

\medskip
 
The evaluation of $T_{a,atm}$ can be obtained through direct measurement and by 
calculing from the profile ($p$, $T$, $u$). The accuracy of the two methodes is 
comparable between 2 and 5 GHz, whilst at higher frequencies is preferable 
a direct measurement (Ajello et al. 1995). We describe here the method of 
the direct measurement since in the works discussed here this method is used. 
In particular, we follow the review of Bersanelli et al. (1995) 
 that 
measure the atmospheric emission at many frequencies (1.47, 2.0, 3.8, 7.5, 
10 and 90 GHz) in three campaigns of White Mountain and two campaigns 
of South Pole. 
 
For a nonscattering and nonrefrancting atmosphere in thermal equilibrium, 
the equation of transfer for a signal from outside the atmosfhere in units 
of antenna temperature can be written (Waters 1976) 
 
\begin{eqnarray}\label{eqn:Tobs}
T(H_{obs}) & = & T_{\infty}{\rm exp}[-\tau_{\nu} (0,s_{obs})]+ \nonumber \\
	& & \mbox{}+\int_{0}^{s_{obs}} T_{phys}(s^{\prime}){\rm exp}[-\tau_{\nu} (s^{\prime}-s_{obs})]k_{\nu}(s^{\prime})ds^{\prime}, 
\end{eqnarray}

\noindent
where $T_{\infty}$ is the antenna temperature of the background signal (in this case 
emission from the Galaxy, extragalactic sources and the CMB), $H_{obs}$ is the 
altitude of the observing site, $T_{phys}=c^2B_{\nu}(T)/2k_B\nu^2$  with $B_{\nu}(T)$ the Planck function and $k_B$ 
Boltzmann's constant [$T_{phys}(s^{\prime})$ approaches the physical temperature at $s^{\prime}$ in the 
Rayleigh-Jeans approximation] and $\tau_{\nu}$ and $k_{\nu}$ are, respectively, the optical 
depth and volume attenuation coefficient at frequency $\nu$. The second term in 
equation (\ref{eqn:Tobs}) is the atmospheric antenna temperature $T_{a,atm}$. 
 
At low frequencies $T_{a,atm}$ is dominated by emission from the $O_2$ continuum, 
characterized by the attenuation coefficient (Danese \& Partridge 1989): 

\begin{equation}\label{eqn:1}
(k_{\nu})_{O_{2,c}}=\left(1.12\mbox{x}10^{-4}p\theta^2\frac{\gamma_0}{\gamma_0^2+\nu^2}+3.49\mbox{x}10^{-11}p^2\theta^{2.5}\right)\nu^2\;\;{\rm dB Km}^{-1},
\end{equation}
 
\noindent
where $\theta=300/T$ is the relative inverse temperature parameter in K, $p$ is the dry 
air pressure in kPa and $\gamma_0$ is the width parameter for the $O_2$ continuum 
in GHz. The latter can be represented as follows: 

\begin{equation}\label{eqn:2}
\gamma_0=a(p+1.1e)\theta^b,
\end{equation}
 
\noindent
where $e$ is the partial water vapor pressure in kPa; the parameters $a$ in 
GHz  kPa$^{-1}$ and $b$, adimensional, determine through equations (\ref{eqn:1}) and (\ref{eqn:2})
the amplitude of the $O_2$ continuum emission. 
 
At frequencies above  $\sim$10 GHz, the contribution of water vapor becomes 
important. In the limit $\tau_{\nu}\ll1$ from the second term on the right-hand side 
of equation (\ref{eqn:Tobs}), the atmospheric temperature can be expressed as (Partridge 
et al. 1984) 

\begin{equation}\label{eqn:3}
T_{a,atm}(\nu)=T_{a,O_2}(\nu)+wT_{a,H_2O}(\nu),
\end{equation}
 
\noindent
where $w$ is the precipitable water vapor content in millimeters and the 
components $T_{a,O_2}$ and $T_{a,H_2O}$ are expressed in K and Kmm$^{-1}$, respectively. 
The quantity $w$ is rapidly variable. Simultaneous measurements at two 
frequencies $\nu_1$ and $\nu_2$ can be used to test the model. 

\medskip
 
The antenna temperature of the atmosphere, $T_{a,atm}$, is measured with zenith 
scans in which each radiometer repeatedly observed the sky at a set of 
zenith angles, $Z$. In general, observations at two angles $Z_1$ and $Z_2$ 
produce an estimate of $T_{a,atm}$. The measured signal difference, $\Delta S_{Z_1/Z_2}$, is 
related to the atmospheric antenna temperature at zenith by: 

\begin{equation}
T_{a,atm,Z_1/Z_2}(0)\simeq\frac{G\Delta S_{Z_1/Z_2}-\Delta T_{gal,Z_1/Z_2}-\delta T_{Z_1/Z_2}}{\langle f(Z_1) \rangle - \langle f(Z_2) \rangle}\langle f(0) \rangle,
\end{equation}
 
\noindent
where $G$ is the radiometer gain constant, repeatedly measured during the 
experiment, and 

\begin{equation}\label{fz}
\langle f(Z) \rangle = \int f(Z;\theta,\phi)g(\theta,\phi)d\Omega
\end{equation}
 
\noindent
is the convolution of the atmospheric air mass, $f(Z;\theta,\phi)$, with the antenna gain 
pattern, $g(\theta,\phi)$, when pointing the radiometer at a zenith angle $Z$. The term 
$f(Z)$ must take into account the Earth's curvature 

$$f(Z)=\frac{1+r}{({\rm cos}^2Z+2r+r^2)^{1/2}},$$
 
\noindent
where $r$ is the atmospheric height in units of Earth's radius. For small 
$Z$, $f(Z)\simeq {\rm sec}(Z)$. 
 
The quantity $\Delta T_{gal,Z_1/Z_2}$ is the differential Galactic background component, 
which is subtracted based on low-frequency surveys (Haslam et al. 1982). The 
term $\delta T_{Z_1/Z_2}$ is the resultant of second-order, differential corrections due 
to ground radiation $\delta T_{gr,Z_1/Z_2}$, angle-correlated instrumental effects $\delta T_{inst,Z_1/Z_2}$, and possible 
radio-frequency interference (RFI) $\delta T_{RFI,Z_1/Z_2}$: 

\begin{equation}
\delta T_{Z_1/Z_2}=\delta T_{gr,Z_1/Z_2}+\delta T_{inst,Z_1/Z_2}+\delta T_{RFI,Z_1/Z_2}+...
\end{equation}
 
\noindent
The relative importance of the various sources of uncertainty depends on 
frequency and site. In particular, at low frequencies the uncertainty is 
dominated by the subtraction of the large Galactic background through the 
propagation of the errors of the 408 MHz map (Haslam et al. 1982). In Table \ref{tab:error} 
we reporte the values of the main sources of systematic error and an 
estimation of the statistical error on the measure of $T_{a,atm}$. 

\medskip

\begin{table}[htbp]
\tiny
\begin{center}
\begin{tabular}{ccccccccc}
\hline
\hline
Frequency & Galaxy & Ground & Instruments & Gain & Pointing & Beam & Total & Statistical \\
(GHz) & (K)) & (K) & (K) & (K) & (K) & Pattern (K) & Systematic (K) & (K) \\
\hline
1.47 & 0.171 & 0.014 & 0.047$^b$ & 0.006 & 0.028 & 0.032 & 0.193$^c$ & 0.027 \\
2.0 & 0.081 & 0.014 & 0.035 & 0.006 & 0.014 & 0.026 & 0.094 & 0.014 \\
3.8 & 0.023 & 0.035 & 0.029 & 0.020 & 0.005 & 0.019 & 0.058 & 0.004 \\
7.5 & 0.010 & 0.041 & 0.013 & 0.004 & 0.007 & 0.014 & 0.047 & 0.037\\
10 & 0.019 & 0.025 & ... & 0.012 & 0.006 & 0.025 & 0.041 & 0.034 \\
90 & $<0.001$ & 0.030 & 0.009 & 0.097 & 0.057 & 0.006 & 0.117 & 0.200 \\
\hline
\end{tabular}
\end{center}
$^a$ The reported values are for pencil beam; at $\nu < 7.5$ GHz from South Pole, the others from White Moutains

$^b$ Extrapolated for a vertical flip set with angle dependence

$^c$ An error of 0.061 K for the RFI emission is considerated
\caption{Values of the main sources of systematic error and an estimation of the statistical error on the measure of $T_{a,atm}$}
\label{tab:error}
\end{table}

\medskip

\noindent
From the comparation with the model of Danese \& Partridge (1989), above 
described, we note a general consistence between the experimental and the 
theorical results. 
 
We report here also the recent result of Mandolesi et al. (1988) who measure 
the atmospheric temperature at 94 GHz with a balloon, obtaining a value 
of $T_{a,atm} \simeq $~0.015$\pm$0.006~K at an altitude of $\simeq$~38~Km,
in strict agreement with the atmospheric emission models.
  
\bigskip

\section{Range 10-37 GHz}\label{sez:10-37}

In the range between 10 and 37 GHz (see Table \ref{tab:10-37}) we can distinguish two kinds of 
observations . In the first case 
(Kogut et al. 1988 and De Amici et al. 1985) the observations are taken from the 
ground with instruments like in observations at low frequencies. For these, we 
refer to the subsection \ref{sez:2-9} and previous ones where an accurate analysis 
of problems and errors of this kind of experiments is already given. 
Other recent observations (Staggs et al. 1996b; Johnson \& Wilkinson 
1987) are carried out with balloons. Flying at the altitude of about 25 km,  
the problems due to the atmosphere, that dominate the uncertainty of the 
ground measures, are reduced. 

\medskip

\begin{table}[htbp]
\begin{center}
\begin{tabular}{llcl}
\hline
\hline
\multicolumn{1}{c}{$\nu$}  &  \multicolumn{1}{c}{$\lambda$}& $T^{th}_{CMB}$ & Reference \\
\multicolumn{1}{c}{(GHz)}  &  \multicolumn{1}{c}{(cm)} & (K) & \\
\hline
10 & 3.0 & $2.62\pm0.058$ & Kogut et al. 1988, Ap.J., 325, 1 \\
10.7 & 2.8 & $2.730\pm0.014$ & Staggs et al 1996b, Ap.J., 473, L1 \\
19.0 & 1.58 & $2.78^{+0.12}_{-0.17}$ & Stokes et al. 1967, Phys. Rev. Lett., 19, 1199 \\
20 & 1.5 & $2.0\pm0.4$ & Welch et al 1967, Phys. Rev. Lett., 18, 1068 \\
24.8 & 1.2 & $2.783\pm0.089$ & Johnson \& Wilkinson 1987, Ap.J. Let , 313, L1 \\
31.5 & 0.95 & $2.83\pm0.07$ & Kogut et al. 1996, Ap.J., 470, 653 \\
32.5 & 0.924 & $3.16\pm0.26$ & Ewing et al 1967, Phys. Rev. Lett., 19, 1251 \\
33.0 & 0.909 & $2.81\pm0.12$ & De Amici et al. 1985, Ap. J.,298, 710 \\
35.0 & 0.856 & $2.56^{+0.17}_{-0.22}$ & Wilkinson, 1967, Phys. Rev. Lett., 19, 1195 \\
37 & 0.82 & $2.9\pm0.7$ & Puzanov et al., 1968, Sov. Astr., 11, 905 \\
\hline
\end{tabular}
\end{center}
\caption{Values of the CMB temperature at $10\leq\nu\leq 37$ GHz}
\label{tab:10-37}
\end{table}

\medskip
 
We analyse here the problems of the balloon experiments. The antenna 
temperature of the CMB comes from (Johnson \&Wilkinson 1987): 

\begin{eqnarray}\label{eqn:Tpallone}
T_{a,CMB} & = & G(S_a-S_{load})+T_{a,load}-T_{a,offset}-T_{a,WND}-T_{a,HRN}- \nonumber\\
          &   & \mbox{}-T_{a,for}- T_{a,gr}-T_{a,IGS},
\end{eqnarray}
 
\noindent
where $T_{a,WND}$ and $T_{a,HRN}$ are the contributions from the window and the horn 
respectively, $T_{a,for}$ is the sum of the atmosphere contribution and the dipole 
contribution and $T_{a,IGS}$ is the screen contribution. All other terms are already 
described in the previous subsections. In Table \ref{tab:10-37} we find the values of 
the terms in equation (\ref{eqn:Tpallone}) and the relative uncertainties in the two  
frequency bands of the experiments here considered. 

\medskip 

\begin{table}[htbp]
\begin{center}
\begin{tabular}{lcc}
\hline
\hline
 & \multicolumn{2}{c}{$\nu$ (GHz)}\\
\cline{2-3}
Temperature (K) & 10.7 & 24.8 \\
\hline
$G(S_a-S_{load})$ & $-0.011\pm0.001$ & \\
Source ($T_{a,load}$) & $2.749\pm0.004$ & $2.870\pm0.008$ \\
Offset ($T_{a,offset}$) & $0.007\pm0.007$ & ... \\
Horn ($T_{a,HRN}$) & $0.007^{+0.003}_{-0.003}$ & $0.050\pm0.012$ \\
Window ($T_{a,WND}$) & $0.001\pm0.001$ & $0.036\pm0.012$ \\
Screens ($T_{a,IGS}$) & $0.003\pm0.003$ & ... \\
Ground ($T_{a,gr}$) & $>0.005$ & $0.004\pm0.004$ \\
Atmosphere ($T_{a,atm}$) & $0.003\pm0.003$ & $0.002\pm0.002$ \\
Galaxy ($T_{a,gal}$) &  & $0.001\pm0.001$ \\
Dipole ($T_{a,dip}$) & $0.002\pm0.002$ & $-0.002\pm0.001$ \\
$T_{CMB}^{th}$ & $2.730\pm0.014$ & $2.783\pm0.089^a$ \\
Reference & Staggs et al. 1996b & Johnson \& Wilkinson 1986 \\
\hline
\end{tabular}
\end{center}
\caption{Values and errors from the balloon's experiments}
\label{tab:10-37r}
$^a$ \footnotesize{The error is conservative and comes from adding up the errors} \\
\end{table}

\medskip
 
The error on $T_{CMB}^{th}$ is abundantly dominated by systematic errors. 
Statistical errors are considered and include the radiometer 
noise and the uncertainty on the gain $G$ and on the evaluation of $T_{a,offset}$ 
(Johnson \& Wilkinson 1987). 
 
In the balloon observations the problem of the atmospheric emission is reduced 
when flying at an altitude of about 25 km. Atmospheric emission is 
evaluated by fitting pressure models to the observed decrease in the sky 
temperature during ascent (Johnson \& Wilkinson 1987). In regards to the 
foregrounds emission, the Galactic term is very small at these frequencies. 
This is extrapolated from the continuum data at 408 MHz (Haslam et al. 1982) 
using a spectral index of $2.8\pm0.1$ (Staggs et al. 1996b). The data is 
also corrected for the dipole of the CMB, as observed by COBE (Fixsen et al. 
1996). After the substraction of the atmospheric and foreground effects, 
the increase in sky temperature, due to the change of the observation angle, 
is used to constrain the magnitude of the ground radiation. The largest 
contribution to $T_{a,gr}$ comes from the ground radiation diffracting over the 
front edge of the ground screen. In principle, radiation from the ground 
might reflect off the balloon and into the horn antenna, but since the 
balloon fills only few sr and has small reflectivity, this effect is 
negligible (Staggs et al. 1996b). The screens limit also the entry into 
antenna of the spurious signal, but the ambient temperature of the inner 
ground screen (IGS), which is fixed to the dewar, is near enough to the 
beam to contribute to the signal (Staggs et al. 1996b). The magnitude of 
this effect is determined after the flight by lining the inner surface of 
the inner ground screen with microwave absorber and measuring this lining's 
contribution to the sky signal. These results are scaled to account for the 
much lower emissivity of the aluminum relative to the microwave absorber 
and the reduced effective emissivity of the absorber at low incidence angles. 
The terms $T_{a,HRN}$ and $T_{a,WND}$ take into account the corrugated horn emission and 
the window emission respectively. $T_{a,HRN}$ includes the effect of the short 
waveguide transition coupling the horn to the waveguide switch. 
 
Finally, the value of $T_{a,offset}$ is determined by removing the horn from the dewar and by 
replacing with a second thermally regutaled waveguide load, identical in 
construction to the reference load (Staggs et al. 1996b). If the radiometer 
were ideal, when the measured temperature of the two waveguide loads were 
equal, no differential signal would be observed from the radiometer. Is 
observed a difference of few mK. Since their origin is not yet certain, 
errors equal to the magnitudes of the offsets are assigned to their removal. 
 
\medskip

The resuls due to the balloon observations are in good agreement with the 
value of FIRAS (Mather et al. 1999), that is a $T_{CMB}$ temperature of 
$2.725\pm0.002$ K (95\% CL). 

\bigskip

\section{Above 50 GHz}\label{sez:50-}

We report in Table \ref{tab:50-} the values of the CMB temperature at frequencies 
above 50 GHz obtained from ground or from balloon flights (Bersanelli 
et al. 1990), even if at these frequencies the FIRAS and COBRA's measures 
are more accurate. Note the value of Schuster et al., in good agreement 
with the FIRAS measure (Mather et al. 1999). In the subsection 
1.8 will be given a short description of the experiments based on the
analyse of molecular lines.

\medskip

\begin{table}[htbp]
\begin{center}
\begin{tabular}{llcl}
\hline
\hline
\multicolumn{1}{c}{$\nu$}  &  \multicolumn{1}{c}{$\lambda$}& $T^{th}_{CMB}$ & Reference \\
\multicolumn{1}{c}{(GHz)}  &  \multicolumn{1}{c}{(cm)} & (K) & \\
\hline

53 & 0.57 & $2.71\pm0.03$ & Kogut et al. 1996, Ap.J., 470, 653 \\
83.8 & 0.358 & $2.4\pm0.7$ & Kislyakov et al. 1971, Sov. Ast., 15, 29 \\
90 & 0.33 & $ 2.46^{+0.40}_{-0.44}$ & Boyton et al. 1968, Phys. Rev. Lett., 21, 462 \\
90 & 0.33 & $2.61\pm0.25$ &  Millea et al. 1971, Phys. Rev. Lett., 26, 919 \\
90 & 0.33 & $2.48\pm0.54$ & Boynton \& Stokes 1974, Nature, 247, 528 \\
90 & 0.33 & $2.60\pm0.09$ &  Bersanelli et al. 1989, Ap.J., 339, 632 \\
90 & 0.33 & $2.712\pm0.020$ & Schuster et al. UC Berkeley PhD Thesis \\
90.3 & 0.332 & $<2.97$ &  Bernstein et al. 1990, Ap.J., 362, 107 \\
90 & 0.33 & $2.72\pm0.04$ & Kogut et al. 1996, Ap.J., 470, 653 \\
154.8 & 0.194 & $<3.02$ & Bernstein et al. 1990, Ap.J., 362, 107 \\
195.0 & 0.154 & $<2.91$ & Bernstein et al. 1990, Ap.J., 362, 107 \\
266.4 & 0.113 & $<2.88$ & Bernstein et al. 1990, Ap.J., 362, 107 \\
\hline
\end{tabular}
\end{center}
\caption{Values of the CMB temperature at frequencies $>50$ GHz. The measures from FIRAS, COBRA and CN molecules are excluded}
\label{tab:50-}
\end{table}

\bigskip

\section{Measures based on the analyse of the molecular lines}\label{sez:CN}

Before the COBE and COBRA's measurements (section \ref{sez:satelliti}), molecular absorption 
line measurements provided some of the most precise values for the CMB 
temperature. This method has the interesting aspect to represent one 
of the few methodes able to verify in a direct way the homogeneity of the CMB, i.e.
the scaling law of the background temperature with the redshift $z$. The standard Friedman 
cosmology predicts in fact a simple relationship between the temperature of the 
CMB radiation and the redshift: 

\begin{equation}\label{eqn:t(z)}
T_{CMB}(z)=T_{CMB}(0)(1+z),
\end{equation}
 
\noindent
where $T_{CMB}(0)$ is the CMB temperature today. 
 
The indirect measure of the CMB temperature through the molecular observation 
let's us think that it is possible to determine the CMB temperature also at 
cosmological distancies, that is at redshifts higher than 1. The blackbody 
temperature at higher redshifts can be measured indirectly by using atomic 
fine-structure transitions in absorbers toward high redshift quasars (Bahcall 
\& Wolf 1968). The first attempt gave an upper limit for the CMB temperature, 
$T_{CMB} < 45$ K, at $z=2.309$ from limits on the fine-structure excitation of C II 
toward PHL 957 (Bahcall et al. 1973). The recent observations use the C I, 
because the energy separations in its fine-structure levels are smaller 
comparated to other abundant species (such as O I, C II, Si II, N II). 
The ground term of C I is split into three levels (J=0, 1, 2) with J=0-1 
and J=1-2 separations of 23.6 K and 38.9 K (or 0.61 mm and 0.37 mm). C~II 
is another good species to use for the CMB measurements at high redshifts 
because it has reasonably small energy separation between its fine-structure 
levels, 91.3 K. 
 
There are several difficulties in carrying out measurements of $T_{CMB}(z)$ with 
quasar absorbers. First, the ground state C I absorbtion lines are often 
weak and difficult to detect in quasar absorbers at high redshift. Second, 
other non-cosmological sources such as collisions and pumping by UV 
radiation can also populate the excited fine-structure levels of C I. Thus, 
the excitation temperature derived is an upper limit to the CMB temperature, 
unless the local excitation can be estimated. Third, most 
absorption lines from abundant species such as O I, C II, SI II, N II show 
strong saturation in their ground state transitions and hence the population 
ratio of their excited state to the ground state cannot be accurately 
determined. 
 
\begin{figure}
\includegraphics{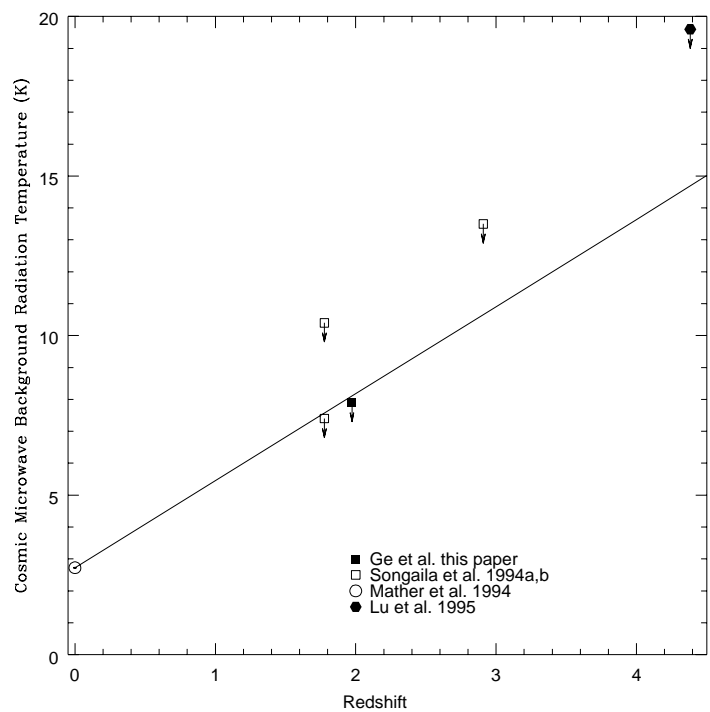}
\caption{Comparation between the measures of the $T_{CMB}(z)$ and the scale-law of equation (\ref{eqn:t(z)})(solid line). Figure from Ge et al. 1995}
\label{fig:CN}  
\end{figure}

The results of the observations using C I and C II are reported in Table \ref{tab:CI}, 
while Figure \ref{fig:CN} shows the comparation between the measures and the scale-law 
of equation (\ref{eqn:t(z)})  (solid line). 

\begin{table}[h]
\begin{center}
\begin{tabular}{lccll}
\hline
\hline
$z$ & $T$ (K) & Molecul & Quasar & Reference \\
\hline
1.776 & $<16 a 2\sigma$ & C I & QSO 1331+170 & Meyer et al. 1986, Ap.J., 308, L37 \\
1.776 & $7.4\pm0.8$ & C I & QSO 1331+170 & Songaila et al. 1994b, Nature, 371, 43 \\
1.9731 & $7.9\pm1.0$ & C I & QSO 0013-004 & Ge et al. 1996, astro-ph/9607145 \\
2.309 & $<45 K a 2\sigma$ &  C II & PHL 957 & Bahcall et al. 1973, Ap.J., 182, L95 \\
2.909 & $<13.5 K a 2\sigma$ & C II & QSO 0636+680 & Songaila et al. 1994, Nature, 368, 599 \\
4.3829 & $<19.6 K a 3\sigma$ & C II & QSO 1202-07 & Lu et al. 1995, Preprint \\
\hline
\end{tabular}
\end{center}
\caption{Values of the CMB temperature due to the observation of the fine-structure transition of the C I and C II}
\label{tab:CI}
\end{table}

\bigskip
 
There are at least three diatomic molecules common in interstellar clouds 
that have low-lying rotational energy states which can be populated by 
the CMB thermal radiation: CN, CH and CH$^+$. We will 
concentrate on the cyanogen, since measurements of the population ratio in the 
other two molecules are very difficult. 

\medskip
 
We will briefly show like it is possible to determine the CMB temperature 
from the observations of the asborption lines of the interstellar molecules 
immersed in the CMB  thermal radiation by following 
the approach of Partridge (1995). Let us begin by considering a collection of 
very simple quantum mechanical systems with just two energy states $E_A$ and 
$E_B$, with $E_B<E_A$; these systems could be atoms or molecules. Place them 
in an oven maintained at temperature $T$ and allow them to reach 
thermodynamic equilibrium. The ratio of the numbers in the two energy states, 
$n_A$ and $n_B$, is then given by the Boltzmann's equation: 

\begin{equation}\label{na/nb}
\frac{n_B}{n_A}=\frac{g_B}{g_A}{\rm exp}[(E_A-E_B)/kT],
\end{equation}
 
\noindent
where $g_A$ and $g_B$ are the so-called statistical weights of the two 
energy states, which can  be calculated from their quantum numbers 
(in the cases we will consider, $g=2J+1$). It is easy to see from equation 
(\ref{na/nb}) that energy state B will be appreciably populated only if $E_B-E_A\;\;\lsim\;\; kT$. 
More to the point, we see that if $g_A$ and $g_B$ are known, $T$ can be 
found directly from a measurement of the population ratio. 

In the molecules 
CN we have an energy difference $E_1-E_0$, corresponding to a wavelength 
$\lambda=hc/(E_1-E_0)=2.64$ mm, and an additional rotational state with $E_2-E_1=2(E_1-E_0)$, corresponding at a 
wavelength of 1.32 mm. Thus, measurements of the population ratios $n_1/n_0$ 
and $n_2/n_1$ permit us to determine $T_0$ at 2.64 and 1.32 mm, respectively. 
The ratios of statistical weights appearing in equation (\ref{na/nb}) are $g_1/g_0=3$ 
and $g_2/g_1=5/3$, respectively. 

\medskip
 
The values of the CMB temperature, obtained from the absorption lines of CN, 
are given in Table \ref{tab:CN}. They provide one of the few means to verify the 
homogeneity of the CMB at least on scales of a few hundred parsecs (Crane 
1995). Clearly a few hundred parsecs is not an interesting scale 
cosmologically, but any variations on these scales would be indicative of 
larger variations on larger scales. It should be possible to measure the 
CMB temperature in the Magellanic clouds (Crane 1995). 

\begin{table}[htbp]
\footnotesize
\begin{center}
\begin{tabular}{llccl}
\hline
\hline
\multicolumn{1}{c}{$\nu$}  &  \multicolumn{1}{c}{$\lambda$}& $T^{th}_{CMB}$ & Observed & Reference \\
\multicolumn{1}{c}{(GHz)}  &  \multicolumn{1}{c}{(cm)} & (K) & star & \\
\hline
113.6 & 0.264 & $2.70\pm0.04$ & z Per & Meyer \& Jura 1985, Ap.J., 297, 119 \\
113.6 & 0.264 & $2.74\pm0.05$ & z Oph & Crane et al. 1986, Ap.J., 309, 822 \\
113.6 & 0.264 & $2.76\pm0.07$ & HD21483 & Meyer et al. 1989, Ap.J. Lett., 343, L1 \\
113.6 & 0.264 & $2.796^{+0.014}_{-0.039}$ & $\zeta$ Oph & Crane et al. 1989, Ap.J., 346, 136 \\
113.6 & 0.264 & $2.75\pm0.04$ & $\zeta$ Per & Kaiser \& Wright 1990, Ap.J. Lett., 356, L1 \\
113.6 & 0.264 & $2.834\pm0.085$ & HD154368 & Palazzi et al. 1990, Ap.J., 357, 14 \\
113.6 & 0.264 & $2.807\pm0.025$ & 16 stars & Palazzi et al. 1992, Ap.J., 398, 53 \\
113.6 & 0.264 & $2.279^{+0.023}_{-0.031}$ & 5 stars & Roth et al. 1993, Ap.J., 413, L67 \\
227.3 & 0.132 & $2.656\pm0.057$ & 5 stars & Roth et al. 1993, Ap.J., 413, L67 \\
227.3 & 0.132 & $2.76\pm0.20$ & $\zeta$ Per & Meyer \& Jura 1985, Ap.J., 297, 119 \\
227.3 & 0.132 & $2.75^{+0.24}_{-0.29}$ & $\zeta$ Oph & Crane et al. 1986, Ap.J., 309, 822 \\         
227.3 & 0.132 & $2.83\pm0.09$ & HD21483 & Meyer et al. 1989, Ap.J. Lett., 343, L1 \\
227.3 & 0.132 & $2.832\pm0.072$ & HD154368 & Palazzi et al. 1990, Ap.J., 357, 14 \\
\hline
\end{tabular}
\end{center}
\caption{Values of the CMB temperature, measured through the molecules CN}
\label{tab:CN}
\end{table}

\medskip

To determine the 
CMB temperature from the population ratios of the rotational states of CN, 
the possibility that other processes (i.e. collisions between electrons and 
the CN molecules) might alter these must be considered 
(i.e. collisions between electrons and the CN molecules), since these can 
lead us to overestimate $T_0$ using equation 1.25 with no correction for such 
collisional excitation (Partridge 1995). Unfortunately, the magnitude of the 
required correction depends on the electron density in the interstellar 
cloud, $n_e$, and this quantity can vary from cloud to cloud. The values of 
the correction for the most observed stars are given in Table \ref{tab:Tloc} (Roth et al. 1993). 

\begin{table}[h]
\begin{center}
\begin{tabular}{lc}
\hline
\hline
Star & $T_{loc}$ (K) \\
\hline
$\zeta$ Ophiuchi & $0.0^{+0.031}_{-0.0}$ \\
$\zeta$ Persei & $0.0^{+0.031}_{-0.0}$ \\
HD 21483 & $0.075\pm0.018$ \\
HD 27778 & $0.020\pm0.020$ \\
HD 154368 & $0.035\pm0.009$ \\
\hline
\end{tabular}
\end{center}
\caption{Values of the correction due to local excitation for the most observed stars (Roth et al. 1993)}
\label{tab:Tloc}
\end{table}

\bigskip

\section{Measures from satellite and rocket}\label{sez:satelliti}

\bigskip

\subsection{FIRAS}\label{sez:FIRAS}

The FIRAS (Far Infrared Absolute Spectrophotometer) instrument aboard the 
COBE (COsmic Background Explorer) satellite was designed to measure the 
spectrum of the CMB temperature. 
 
The FIRAS is a polarizing Michelson interferometer. It measures the spectral 
difference between a $7^{\circ}$ patch of sky and an internal blackbody. The symmetric 
FIRAS optics are differential, with two input and two output ports. One input 
port receives emission from the sky, defined by a non-imaging concentrator. 
The other input port receives emission from an internal reference calibrator 
(emissivity $\simeq0.98$) with an associated concentrator. Each of the two output 
beams is split by a dichroic filter into low and high frequency beams, 
separated at 20 cm$^{-1}$, feeding four silicon composite bolometer detectors 
operated simultaneously. An external blackbody calibrator provides the 
critical absolute calibration. During calibration the sky aperture is 
completely filled by the external calibrator with an emissivity $>0.99997$, 
calculated and measured. The external calibrator is isothermal to better 
than 1 mK at 2.7 K according to calculation. The spectrum uncertainty due to 
the calibrator is approximately 10 parts per million. 
 
The temperature of the two calibrators and associated concentrators are 
controllable from 2 to 25 K. Redundant thermometers measure the temperatures 
of these four temperature controlled elements and other infrared emitters 
such as the moving mirrors, the mechanical structure, and the detector 
housings. When observing the sky, the spectrometer is operated with its 
output nearly nulled, by adjusting the internal calibrator temperature. This 
reduce sensitivily the gain errors and instrument drifts. 

\bigskip

A first analysis of the FIRAS data was considered by Mather et al. (1990) for
measures in the range $1<\nu<20$ cm$^{-1}$. The error bars are a conservative 
valutation of the systematic errors in the calibration algorithm, corresponding
to 1\% of the spectrum peak intensity. Mather et al. (1990) obtain a temperature
of $2.735\pm0.060$ K in the range above mentioned.

A more refined analysis of the FIRAS data in the range $2<\nu<20$ cm$^{-1}$
was made by Fixsen et al. (1994) and Fixsen et al. (1996), improving the 
calibration. The bias of some pixels is corrected, the effects of the Cosmic
Rays hits on the dectector are removed, data with a large number of glitches
are deweighted relative to data with few glitches, 320 points in the spectrum 
are used rather than 256 and the data are destriped after the calibration.
Using the calibration method by Fixsen et al. (1994), Mather et al. (1994) 
derived the absolute temperature of the CMB as $2.726\pm0.010$ K (95 \% CL), with
a conservative systematic uncertainty estimate. They note a discrepancy 
between the thermometers and the color temperature scale.

A subsequent analysis of FIRAS data has been performed by 
Fixsen et al. (1996), that obtained 
a scale temperature of the CMB as $2.728\pm0.004$ (95\% CL). They used three
ways to determine the CMB temperature from the FIRAS data set:

\begin{enumerate}
\item one use the preflight calibration of the external calibrator thermometers, which 
should be good to the nominal 1 mK accuracy of the calibration specification. This 
method gives a CMB temperature of $2.730\pm0.001$ K, with the error entirely dominated by 
the absolute thermometry calibration error on the external calibrator;
\item one uses the data to determine the temperature scale. Ones notes 
the possibility that the high and the low frequency calibration need not agree. The 0.03\% 
frequency uncertainty implies a temperature uncertainty of 0.82 mK. There is an additional 0.3 mK 
error in determining the color temperature once the frequency scale is set, but as this adds in 
quadrature it is negligible. The result of this analysis is that the CMB temperature is $2.7255\pm0.0009$ K;
\item ones can also use the CMB itself. If one assumes the dipole is a result of a Doppler shift 
the shape of the differential spectrum should be $dB_{\nu}/dT$, where $B_{\nu}(T)$ is the Planck function. 
The best fit temperature to the dipole spectrum is a CMB temperature of $2.717\pm0.007$ K. 
The uncertainty is dominated by the uncertainty in fitting the Galaxy radiation which modulates the 
dipole signal which is only 0.1\% of the CMB signal in the Rayleigh-Jeans region.
\end{enumerate}

These three method give results in agreement one each other within three sigma. 
By combining them, the absolute
temperature of the CMB is then $2.728\pm0.004$ K (95\% CL), entirely dominated
by the systematic errors. While this is not a true statistical uncertainty, it quantifies
the uncertainty in the result.

The calibrated destriped sky spectra were then fit to four spatial template.
One uses the FIRAS data to determine the spectra of the four components. The 
data $S(\nu;l,b)$, where $l$, $b$ are Galactic coordinates and $\nu$ is 
frequency, as follows:

\begin{equation}\label{contr}
S(\nu;l,b)=I_0(\nu)+D(l,b)d(\nu)+G_1(l,b)g_1(\nu)+G_2(l,b)g_2(\nu),
\end{equation}

\noindent
where the monopole is represented by the spectrum $I_0(\nu)$; the dipole 
variation is represented by the spatial distribution $D(l,b)$ and the spectrum
$d(\nu)$; and the Galactic emission is represented by one or two spatial
distributions $G_k(l,b)$ and the corresponding spectra $g_k(\nu)$. The fits is
made independently at each frequency, only the spatial variation is assumed.

To make this separation the fuctions $D(l,b)$ for the dipole and $G_k(l,b)$
for the Galactic emission must be specified. The dipole is $D(l,b)={\rm cos}(\theta)$,
where $\theta$ is the angle between the observation and the maximum of the 
dipole, $(l,b)=(264.26^{\circ},+48.22^{\circ})$ (Smooth et al. 1992).

Five templates have been considered instead for $G(l,b)$: 

\begin{enumerate}
\item a plane-parallel, ${\rm csc}|b|$, distribution;
\item the spatial distribution of the power received in the high frequency FIRAS channel 
above 25 cm$^{-1}$. This is used under the assumption that the high frequency radiation is well 
correlated to the low frequency Galactic radiation;
\item the COBE DIRBE 240 $\mu$m map, convolved to FIRAS resolution. This 
has the advantage of being totally independent and low noise but could suffer from beam 
convolution errors. The DIRBE resolution is $\sim0.7^{\circ}$ and the FIRAS resolution is $\sim7^{\circ}$;
\item the COBE DIRBE 140 $\mu$m map, convolved to the FIRAS resolution;
\item the COBE DIRBE 100 $\mu$m map, convolved to the FIRAS resolution.
\end{enumerate}

For Galaxy templates 2-5 one uses the normalization  $\langle G(l,b) \rangle_{|b|>60^{\circ}}=1.074$,
the natural normalization of the ${\rm csc}|b|$ model. The only effect of the 
normalization, of course, is to rescale the Galaxy spectrum.

\bigskip

The monopole  $I_0(\nu)$ can be fitted by a Planck blackbody spectrum at the 
temperature $T_0$ and by a component of astrophysical monopole.

$$I_0=B_{\nu}(T_0)+G_0g(\nu),$$

\noindent
where $B_{\nu}(T_0)$ is a blackbody at temperature $T_0$ and $G_0g(\nu)$ 
is a monopole term.
This second component is of astrophysical nature and can not be interpreted
in terms of variations of $T_0$ or of presence of Bose-Einstein 
or comptonization distortions (Fixsen et al. 1996). By comparing the observation
with the model, one obtains for each channel a residue, which, added to the 
scale temperature value of the spectrum, gives the temperature observed by
FIRAS in that channel. The value of the residues, the measured uncertainty and
the astrophysical monopole are reported in Table \ref{ttres}.

\begin{table}[h]
\begin{center}
\begin{tabular}{rrrr}
\hline
\hline
Frequency &  Residual & Uncertainty & Astrophysical\\ 
$cm^{-1}$ & kJy/sr& 1$\sigma$ & monopole\\
\hline
  2.27&      5& 14&  4\\ 
  2.72&      9& 19&  3\\ 
  3.18&     15& 25& -1\\
  3.63&      4& 23& -1\\ 
  4.08&     19& 22&  3\\ 
  4.54&    -30& 21&  6\\ 
  4.99&    -30& 18&  8\\ 
  5.45&    -10& 18&  8\\ 
  5.90&     32& 16& 10\\ 
  6.35&      4& 14& 10\\ 
  6.81&     -2& 13& 12\\ 
  7.26&     13& 12& 20\\ 
  7.71&    -22& 11& 25\\
  8.17&      8& 10& 30\\ 
  8.62&      8& 11& 36\\ 
  9.08&    -21& 12& 41\\ 
  9.53&      9& 14& 46\\ 
  9.98&     12& 16& 57\\ 
 10.44&     11& 18& 65\\ 
 10.89&    -29& 22& 73\\ 
 11.34&    -46& 22& 93\\ 
 11.80&     58& 23& 98\\ 
 12.25&      6& 23&105\\ 
 12.71&     -6& 23&121\\ 
 13.16&      6& 22&135\\ 
 13.61&    -17& 21&147\\ 
 14.07&      6& 20&160\\ 
 14.52&     26& 19&178\\ 
 14.97&    -12& 19&199\\ 
 15.43&    -19& 19&221\\ 
 15.88&      8& 21&227\\ 
 16.34&      7& 23&250\\ 
 16.79&     14& 26&275\\ 
 17.24&    -33& 28&295\\ 
 17.70&      6& 30&312\\ 
 18.15&     26& 32&336\\ 
 18.61&    -26& 33&363\\ 
 19.06&     -6& 35&405\\ 
 19.51&      8& 41&421\\ 
 19.97&     26& 55&435\\ 
 20.42&     57& 88&477\\ 
 20.87&   -116&155&519\\ 
 21.33&   -432&282&573\\
\hline
\end{tabular}
\end{center}
\caption{Monopole spectrum (Fixsen et al. 1996) }
\label{ttres}
\end{table}

In Table \ref{tab:FIRAS} we report the values and the statistical error of the CMB temperature 
for each channel of FIRAS. 

\begin{table}[htbp]
\begin{center}
\begin{tabular}{llcc}
\hline
\hline
\multicolumn{1}{c}{$\nu$}  &\multicolumn{1}{c}{$T^{th}_{CMB}$}&\multicolumn{1}{c}{Upper} & \multicolumn{1}{c}{Lower} \\
\multicolumn{1}{c}{(GHz)}  &\multicolumn{1}{c}{(K)} &\multicolumn{1}{c}{error (K)} &\multicolumn{1}{c}{error (K)} \\
\hline
$    68.1$&$  2.72804$&$   .00011$&$   .00011$\\
$    81.5$&$  2.72805$&$   .00011$&$   .00011$\\
$    95.3$&$  2.72807$&$   .00011$&$   .00011$\\
$   108.8$&$  2.72801$&$   .00009$&$   .00009$\\
$   122.3$&$  2.72806$&$   .00007$&$   .00007$\\
$   136.1$&$  2.72792$&$   .00006$&$   .00006$\\
$   149.6$&$  2.72792$&$   .00005$&$   .00005$\\
$   163.4$&$  2.72798$&$   .00004$&$   .00004$\\
$   176.9$&$  2.72807$&$   .00004$&$   .00004$\\
$   190.4$&$  2.72801$&$   .00003$&$   .00003$\\
$   204.2$&$  2.72800$&$   .00003$&$   .00003$\\
$   217.6$&$  2.72803$&$   .00002$&$   .00002$\\
$   231.1$&$  2.72795$&$   .00002$&$   .00002$\\
$   244.9$&$  2.72802$&$   .00002$&$   .00002$\\
$   258.4$&$  2.72802$&$   .00002$&$   .00002$\\
$   272.2$&$  2.72795$&$   .00003$&$   .00003$\\
$   285.7$&$  2.72802$&$   .00003$&$   .00003$\\
$   299.2$&$  2.72803$&$   .00004$&$   .00004$\\
$   313.0$&$  2.72803$&$   .00005$&$   .00005$\\
$   326.5$&$  2.72792$&$   .00006$&$   .00006$\\
$   340.0$&$  2.72786$&$   .00007$&$   .00007$\\
$   353.8$&$  2.72820$&$   .00008$&$   .00008$\\
$   367.2$&$  2.72802$&$   .00008$&$   .00008$\\
$   381.0$&$  2.72798$&$   .00009$&$   .00009$\\
$   394.5$&$  2.72803$&$   .00010$&$   .00010$\\
$   408.0$&$  2.72792$&$   .00010$&$   .00010$\\
$   421.8$&$  2.72803$&$   .00011$&$   .00011$\\
$   435.3$&$  2.72816$&$   .00012$&$   .00012$\\
$   448.8$&$  2.72792$&$   .00013$&$   .00013$\\
$   462.6$&$  2.72785$&$   .00015$&$   .00015$\\
$   476.1$&$  2.72807$&$   .00019$&$   .00019$\\
$   489.9$&$  2.72807$&$   .00023$&$   .00023$\\
$   503.4$&$  2.72816$&$   .00030$&$   .00030$\\
$   516.8$&$  2.72757$&$   .00037$&$   .00037$\\
$   530.6$&$  2.72809$&$   .00045$&$   .00045$\\
$   544.1$&$  2.72845$&$   .00055$&$   .00055$\\
$   557.9$&$  2.72748$&$   .00066$&$   .00066$\\
$   571.4$&$  2.72786$&$   .00080$&$   .00080$\\
$   584.9$&$  2.72821$&$   .00108$&$   .00108$\\
$   598.7$&$  2.72880$&$   .00168$&$   .00168$\\
$   612.2$&$  2.73002$&$   .00311$&$   .00311$\\
$   625.7$&$  2.72316$&$   .00652$&$   .00652$\\
$   639.5$&$  2.70634$&$   .01468$&$   .01468$\\
\hline
\end{tabular}
\end{center}
\caption{Value and statistical error of the CMB temperature for each FIRAS channel (Nordberg \& Smoot 1998)}
\label{tab:FIRAS}
\end{table}

Table \ref{tab:fit} shows the results of the fit by Fixsen et al. (1996).

\begin{table}[htbp]
\begin{center}
\begin{tabular}{lrrrr}
\hline
\hline
 & Fit & Statistical & Systematic & Final \\
 & Value & uncertainty &uncertainty &uncertainty \\
 & 	 &  (1$\sigma$) &  (1$\sigma$) &  (1$\sigma$)\\
\hline
Galaxy Temp & 13.3 & 0.6 & 0.8 & 1.0 K \\
Dipole Amp & 3.372 & 0.004 & 0.006 & 0.007 mK \\
Dipole Temp & 2717 & 3 & 6 & 7 mK \\
Gal Latitude & 48.26 & 0.11 & 0.10 & 0.15 deg \\
Gal Longitude & 264.14 & 0.14 & 0.06 & 0.15 deg \\
CMBR Temp & 2.728 & 0.00001 & 0.002 & 0.002 K \\
\hline
\end{tabular}
\end{center}
\caption{Fit results (Fixsen et al. 1996)}
\label{tab:fit}
\end{table}

A recent revision of the calibration work has fixed the value of the absolute
temperature at $2.725\pm0.002$ (95\% CL) (Mather et al. 1999).

\bigskip

\subsection{COBRA}

The instrument COBRA developed and tested during the 1980s, was launched on 20
January 1990. In launch condition prior to liftoff the skytelescope port is 
closed by a vacuum door, the inside being at a temperature about 20 K. During 
the ascent this door was opened at 150-km altitude allowing the observation of
deep space; it was closed at 100 km on descent. The apogee was 250 km. In this
section we report the analysis of Gush et al. (1990).

COBRA has measured the CMB temperature at  $3<\nu<16$ cm$^{-1}$ ($90<\nu<500$ GHz).
In Table \ref{tab:COBRA} the values and the errors for each channel are reported,
following the Nordberg \& Smoot (1998) report.

\begin{table}[htbp]
\begin{center}
\begin{tabular}{llcc}
\hline
\hline
\multicolumn{1}{c}{$\nu$}  &\multicolumn{1}{c}{$T^{th}_{CMB}$}&\multicolumn{1}{c}{Upper} & \multicolumn{1}{c}{Lower} \\
\multicolumn{1}{c}{(GHz)}  &\multicolumn{1}{c}{(K)} &\multicolumn{1}{c}{error (K)} &\multicolumn{1}{c}{error (K)} \\
\hline
$    54.6$&$    2.789$&$     .100$&$     .100$\\
$    68.1$&$    2.648$&$     .100$&$     .100$\\
$    81.8$&$    2.664$&$     .100$&$     .100$\\
$    95.3$&$    2.737$&$     .010$&$     .010$\\
$   109.1$&$    2.718$&$     .010$&$     .010$\\
$   122.6$&$    2.724$&$     .010$&$     .010$\\
$   136.4$&$    2.753$&$     .010$&$     .010$\\
$   149.9$&$    2.736$&$     .010$&$     .010$\\
$   163.7$&$    2.724$&$     .010$&$     .010$\\
$   177.2$&$    2.735$&$     .010$&$     .010$\\
$   191.0$&$    2.731$&$     .010$&$     .010$\\
$   204.5$&$    2.725$&$     .010$&$     .010$\\
$   218.2$&$    2.734$&$     .010$&$     .010$\\
$   231.7$&$    2.737$&$     .010$&$     .010$\\
$   245.5$&$    2.735$&$     .010$&$     .010$\\
$   259.0$&$    2.733$&$     .010$&$     .010$\\
$   272.8$&$    2.733$&$     .010$&$     .010$\\
$   286.3$&$    2.735$&$     .010$&$     .010$\\
$   300.1$&$    2.735$&$     .010$&$     .010$\\
$   313.6$&$    2.742$&$     .010$&$     .010$\\
$   327.4$&$    2.754$&$     .010$&$     .010$\\
$   340.9$&$    2.743$&$     .010$&$     .010$\\
$   354.7$&$    2.734$&$     .010$&$     .010$\\
$   368.1$&$    2.751$&$     .010$&$     .010$\\
$   381.9$&$    2.752$&$     .010$&$     .010$\\
$   395.4$&$    2.739$&$     .010$&$     .010$\\
$   409.2$&$    2.752$&$     .010$&$     .010$\\
$   422.7$&$    2.772$&$     .010$&$     .010$\\
$   436.5$&$    2.747$&$     .010$&$     .010$\\
$   450.0$&$    2.755$&$     .010$&$     .010$\\
$   463.8$&$    2.762$&$     .010$&$     .010$\\
$   477.3$&$    2.686$&$     .100$&$     .100$\\
$   491.1$&$    2.637$&$     .100$&$     .100$\\
$   504.6$&$    2.683$&$     .100$&$     .100$\\
$   518.3$&$    2.732$&$     .100$&$     .100$\\
$   531.8$&$    2.713$&$     .100$&$     .100$\\
$   545.6$&$    2.719$&$     .100$&$     .100$\\
$   559.1$&$    2.743$&$     .100$&$     .100$\\
$   572.9$&$    2.717$&$     .100$&$     .100$\\
$   586.4$&$    2.723$&$     .100$&$     .100$\\
\hline
\end{tabular}
\end{center}
\caption{Value and error of the CMB temperature for each COBRA channel (Nordberg \& Smoot 1998)}
\label{tab:COBRA}
\end{table}

\medskip

The idea of the experiment is to compare the sky radiation with the radiation
of a  blackbody calibrator. The radiation of the sky is directed to one side of
a differential polarizing two-beam interferometer of symmetric construction;
the second side of the interferometer is illuminated by an identical telescope,
terminated, however, by a conical blackbody calibrator. If the spectrum of the 
sky were thermal, and the temperature matched that of the calibrator, null
interferograms would be produced; deviations from a null interferogram would
show immediately that distorsions of the CMB are present. The calibration of
the internal blackbody was made after the launch by filling the sky telescope
aperture by an independent blackbody whose emissivity is calculated to be more
than 0.999 and whose temperature could be set anywhere in the range 1.8-4.2 K
independently of the temperature of the instrument. The discrepancy of the 
two calibrators is lower than $\pm5$ mK. This is a limit for the accuracy of the
measures.

In the frequency region $3<\nu<16$ cm$^{-1}$,where the signal-to-noise 
ratio is high, the rms scatter of the temperature is 9.5 mK, about 0.33\%
of the mean. The standard deviation of the mean temperature equals 2 mK. 
Outside of this band the confidence of the temperature measurement is not 
good. Nevertheless, one can say with considerable certainty that for 
$20<\nu<30$ cm$^{-1}$ the temperature is less than 3 K.

Various factors contribute to uncertainty in the measured value for the mean 
temperature:
1) the internal calibrator non uniformity [5 mK]; 2) the uncertainties in offset
corrections [3 mK]; 3) the possibile non cancellation of microphonics which 
developed during liftoff [2 mK]; 4) the statistical fluctuations in the measured
spectrum [2 mK] and 5) the uncertainties due to heating of the sky telescope 
by temporary coupling to the warm door during liftoff [5 mK].

\medskip

The mean value of the CMB temperature is also of $2.736\pm0.017$, where the
error is the linear sum of these uncertainties.
This value is in good agreement with the measure of FIRAS in the same range
(see section \ref{sez:FIRAS})

\medskip 
 
\section{Our database}

Through the analysis of the observing data of the CMB temperature measurements, 
already presented above, we have built a reasoned database of these values. 
 
The main purpose of this work is to create a complete collection of the 
measurements easily accessible (Salvaterra 2000). The database is thus an ASCII file. 
This simple format permits a reading of the data through any text editor or 
through programs for data handling.

\begin{figure}
\includegraphics{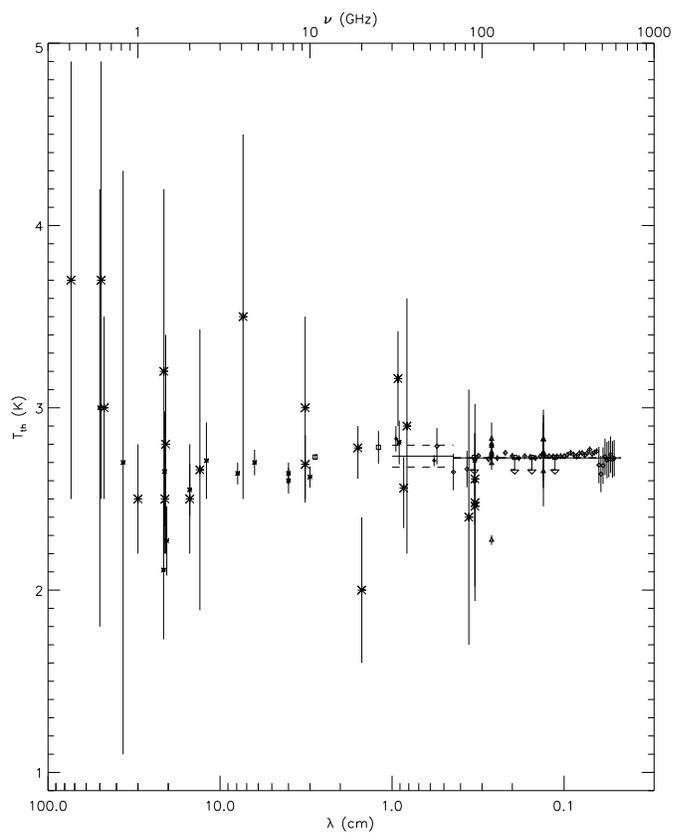}
\caption{Plot of all measures of the CMB thermodynamic temperature. The big stars are the early ground 
measurements, the small stars the recent ground measurements, the squares the balloon measurements, 
the triangles the experiments with the CN molecural and the diamonds the COBRA data. 
The FIRAS data are plot as a solid line.}
\label{fig:tutti}  
\end{figure}

\begin{figure}
\includegraphics{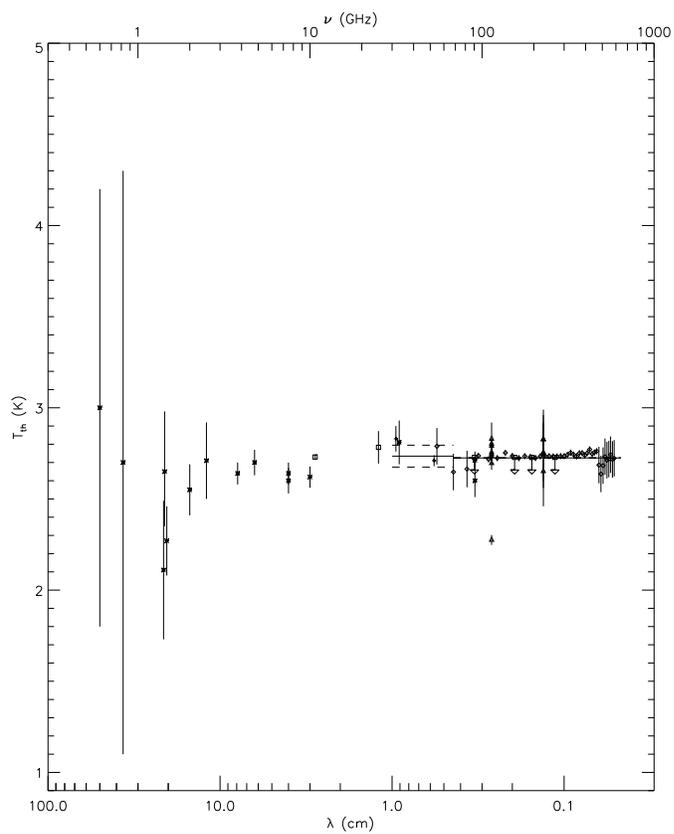}
\caption{Plot of the more recent measures of the CMB thermodynamic temperature. The small stars the 
recent ground measurements, the squares the balloon measurements, the triangles the experiments 
with the CN molecural and the diamonds the COBRA data. The FIRAS data are plot as a solid line.}
\label{fig:rec}  
\end{figure}

\begin{figure}
\includegraphics{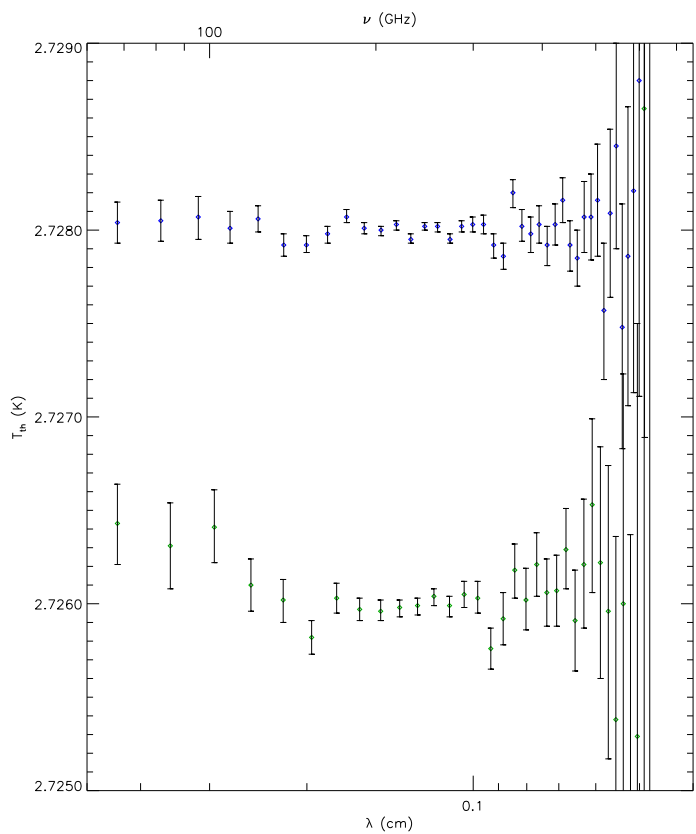}
\caption{A comparation between the FIRAS data reported by Mather et al. (1994) 
and the more refined analisis of these by Fixsen et al. (1996)}
\label{fig:firas}  
\end{figure}

In Figs. \ref{fig:tutti}-\ref{fig:firas} we give some plots of the data. Fig.
\ref{fig:tutti} shows all measures of the CMB thermodynamic temperature, while
in Fig. \ref{fig:rec} reports the more recent measurements. In Fig.~\ref{fig:firas}
we show a comparison between the FIRAS data reported by Mather et al. (1994) 
and the more refined analysis of them by Fixsen et al. (1996).

\subsection{The database structure} 
 
The database is organized in two distinguished files to make simpler the reading: 
either in direct mode through a terminal or through a program. We have 
grouped in the same file all the interesting numerical values so that the reading 
program can extract them easily without the need for  character  
handling. The second file contains instead varied kinds of notes, not so  
interesting under the profile of fits or plots production, but necessary to  
understand the observation problems and to recover other informations. 
Each data is represented by a catalogue number so that we may read the 
two files like a single database and we can at any moment associate a value 
to its reference and notes. The division in two files is therefore motivated 
only by a reason of simplicity and the database must be considered as a single 
object. 
 
The values are arranged in the database in increasing order of frequency (or 
in decreasing wavelength). We prefer this order instead a different layout (i.e. 
in chronological order) so that the detailed division in frequency ranges 
given in the above sections, becomes very clear. 
As far as the measurements with the same observing frequency,  these are 
in chronological order. The FIRAS and COBRA data are, instead, positioned at end of the 
database: they are organized in two separate blocks with an increasing  
frequency order. This way, the access is very easy.

Horizontally  the database is divided in 18 columns, 10 in the first file  
with the numerical values and 8 in the second file with notes and references. 
The first column of both files contains the catalogue number, necessary to  
determine univocally the data. 
 
\medskip 
 
So, the numerical values file is divided in 10 columns. 
 
The first one contains the catalogue number; in the second and third we have  
respectively the observing frequency and wavelength. In the next one the 
measured value of the CMB thermodynamic temperature is given, while in the 5th e 6th column 
we report the absolute values of the upper and lower error associated with 
this measure. 
 
The last four columns deserve a deeper analysis. They are the synthesis  
of the analysis of the main error sources studied in the previous sections. 
They present in order the total amount of the  
statistical uncertainty on the measured value, the amounts related to the two 
main sources of systematic uncertainty and a flag. Obviously, we have these 
values only for the works analysed in detail.
In all other cases, the database is filled in by assigning the entire 
error on the measure to statistical sources and zero value to the systematic 
errors. The last column is a flag that gives an idea of the type of analysis 
we have made in the measurement. So that we can distinguish clearly  
accurately, partially or not at all analyzed data. In Table \ref{tab:flag} 
is reported the meaning of the flag value.

\begin{table}[htbp] 
\begin{center} 
\begin{tabular}{cl} 
\hline 
\hline 
Value & Meaning \\ 
\hline 
0 & the statistical error is on the total\\ 
1 & the statistical error is on one of the terms that give the\\ 
  & total error. This is also a lower limit for this value \\ 
2 & the statistical error is extrapolated from the values of \\ 
  & measurements in close frequencies \\ 
3 & the statistical error is not known \\ 
4 & we have no information on the errors. We assign the entire \\ 
  & error to statistical source\\ 
5 & COBRA problematical data: the total error is conservative. \\ 
  & In the columns are given the errors reported by Gush et al. (1990) \\ 
6 & FIRAS data from the first calibration (Mather et al. 1990) \\
\hline 
\end{tabular} 
\end{center} 
\caption{Meaning of the flag value} 
\label{tab:flag} 
\end{table} 
 
\medskip

The second file presents in its eight columns useful information for the understanding 
of the values that we give in the first part of the database. The first column 
permits to identify the data with its catalogue number. The second and the 
third column give the cause of the two main sources of systematic error that we have 
analyzed in the first part of the database, showing straightaway which kind of  
contributes are the major causes of the uncertainty on the measure. If an 
analysis of the work has not been made, these columns will be obviously 
empty. In the 4th column is reported the year of the observation. 
The 5th column  gives the observing site because 
we have realized that the choice of location is very important for reducing 
the value of the error bars. When the exact site is not known the only given 
indication is whether 
the measure is taken from ground or from balloon. Another important 
indication is in the 6th column that shows which kind of calibrator is 
used in the measurement. For the CN experiments we report in this column 
the observed star or the number of the stars used in the measurement.  
 
The presence of a 'N' in the 7th column refers to a larger note contained in another file. 
In this istance, the catalogue number permits to find immediately 
the note desired. This report contains what was not possible to include 
in the database, but anyway very useful. In this file is also presented 
the flag legend. 
 
The 8th column reports the reference of the published work, where  further 
informations on the experiment can be found.

\subsection{The program of data handling} 
 
We have the requirement to generate different files with different choices of  
data for the fit programs. A simple program is also necessary to 
build through many options the desired data files.   
We have implemented it in FORTRAN. 
 
To start, the user must give the name of the output file. 
Then, other inputs, necessary 
to exploit the FIRAS data, are required.
The program  
asks the FIRAS scala temperature and its uncertainty   
reminding the user anyway the latest value of Mather et al. (1999). 
So, a possible
improved FIRAS calibration can be easily 
taken into account in the future by setting the appropriate value. 
 
The program presents then the main menu of the options. Through these, the 
user can build the output file by considering the required temperature values.
The options are expandable when necessary. 
 
The first 8 options follow the above division in frequency ranges. It is also 
possible to transfer into the output file, as indipendent blocks, the values 
obtained in the various analyzed 
ranges:  0.408-1 GHz, 1-2 GHz, 2.3-9.4 GHz, 10-37 GHz, over 50 GHz (included 
the measures from the study of the CN molecular), the COBRA data, the FIRAS 
data. The eight option trasfers all data. 
The FIRAS data are available with different options, described in the following.
For all the others, the program permits to choose whether the 
data must be written with the total error or only with the statistical error. 
Then, the program writes the data in the output file and returns at the main 
menu so that another choice is possible. 
 
The FIRAS data are available with or without the astrophysical monopole. 
In each case the user can have the data with the total error or with the 
statistical error. The total error comes by adding in quadrature the 
statistical error of the single channel with the error on the scala temperature
given at the begin of the program by the user. 
Then, the program writes the data in the output file and returns at the main 
menu so that another choice is possible.

The next option permits to decide whether to insert or not in the output 
file each singular data. 
For each data is given a short informative table that reports the catalogue 
number, the observation frequency, the measured value of the CMB thermodynamic 
temperature, the upper and the lower error totals and the statistical error. 
Furthermore, it is indicated when the data considered is from the analysis of the 
CN lines, from COBRA or from FIRAS, and the estimate of the statistical 
error.
 
The program asks whether to insert the data 
in the output file and according to the case with which kind of error (total 
or statistical). All available measurements are considered, thereafter 
the program, upon closing, reminds the user the name of the output 
file. Because the run of the data can be long and boring, together with 
the program we give same useful parameter files that automatically 
introduce the answers necessary for building new interesting files. 

In this way, we have a file with the measures of recent ground experiments, 
another one with early ground experiments, a third one from the balloon 
and a fourth one with CN molecular data. 
When using these files, remember that the error associated to the measures 
is the total error. 
 
Another option of the program is to transfer in the output file the  
values obtained in a certain range of years as requested by the user. 
As always, it is possible to have the data either with 
the total error or with the statistical error. Having finished  
writing the data, the program, upon closing, reminds the user the name 
of the output file. 
 
The last option permits to exit the program.  

\medskip

\section{Conclusions}

We have reviewed all the existing measures of the CMB absolute temperature.
After a discussion of the main contamination relevant for the whole set of
observed frequencies, we have considered in detail the different sources
of contamination relevant at the different frequency bands for ground based,
balloon and space observations. We have compared the relative weights of statistical
and systematic uncertainties, by focussing 
in particular on the most recent observations. 
On the basis of this analysis, we have constructed a complete 
and reasoned database of CMB absolute temperatures, that allows to easily 
recognize the relevant informations on the different observations.
The simple database format permits a reading of the data through any text editor or 
through programs for data handling.
We have then implemented a set of tools in FORTRAN, that allow to easily select the 
desired set of measures
and discriminate between the quoted statistical and systematic errors.
This constitutes a the first step for 
a versatile comparison of the existing data with the theoretical predictions for the distorted spectra
in order to derive constraints on physical processes at very high redshifts.

\bigskip
\bigskip

\noindent
{\bf Acknowledgements.} It is a pleasure to thank M.~Bersanelli, L.~Danese,
G.~De~Zotti, N.~Mandolesi and G.~Palumbo for useful and stimulating discussions.

\bigskip
\bigskip

\section{Simbols index}

\noindent
$d(\nu)$ = dipole spectrum; eqn. \ref{contr}

\noindent
$D(l,b)$ = dipole spatial distribution; eqn. \ref{contr}

\noindent
$e^{-\tau}$ = trasmission coefficient of the line; eqn. \ref{eqn:Ta}

\noindent
$E_A$ ($E_B$) = energy in the energy state A (B); eqn. \ref{na/nb}

\noindent
$f(Z;\theta,\phi)$ = atmospheric air mass; eqn. \ref{fz}

\noindent
$g_A$ ($g_B$) = statistical weight of the energy state A (B); eqn. \ref{na/nb}

\noindent
$g_k(\nu)$ = Galactic emission spectrum; eqn. \ref{contr}

\noindent
$g(\theta,\phi)$ = antenna gain pattern; eqn. \ref{fz}

\noindent
$G$ = system gain; eqns. \ref{eqn:Ta}, \ref{eqn:1-2:Ta}, \ref{eqn:2-9:Tacmb}

\noindent
$G_k(l,b)$ = Galactic emission spatial distribution; eqn. \ref{contr}

\noindent
$H_{obs}$ = altitude of the observing site; eqn. \ref{eqn:Tobs}

\noindent
$I_0(\nu)$ = monopole; eqn. \ref{contr}

\noindent
$k_B$ = Boltzmann's constant

\noindent
$k_{\nu}$ = volume attenuation coefficient at frequency $\nu$; eqn. \ref{eqn:Tobs}

\noindent
$(k_{\nu})_{O_{2,c}}$ = attenuation coefficient of the $O_2$; eqn. \ref{eqn:1}

\noindent
$n_A$ ($n_B$) = population in the energy state A (B); eqn. \ref {na/nb}

\noindent
$p$ = prpressure

\noindent
$r$ = power reflection coefficient of the antenna; eqn. \ref{eqn:Ta} 

\noindent
$R_c$ = power reflection coefficient of the cable; eqn. \ref{eqn:teff}

\noindent
$S_a$  = signal from the zenith; eqns. \ref{eqn:Ta}, \ref{eqn:1-2:Ta}, \ref{eqn:2-9:Tacmb}, \ref{eqn:Tpallone}

\noindent
$S_{load}$  = signal from the calibrator; eqns. \ref{eqn:Ta}, \ref{eqn:1-2:Ta}, \ref{eqn:2-9:Tacmb}, \ref{eqn:Tpallone}

\noindent
$S_{syn}$ = synchrotron power

\noindent
$\langle T \rangle$ = averange temperature of the cable between the termination and the receiver; eqn. \ref{eqn:teff}

\noindent
$T_0$ = physical temperature of the system; eqn. \ref{eqn:Ta}

\noindent
$T_a$ = antenna temperature; eqns. \ref{eqn:Ta}, \ref{sky} 

\noindent
$T_{a,atm}$ = atmosphere antenna temperature; eqns. \ref{sky}, \ref{eqn:1-2:Tcmb}, \ref{eqn:2-9:Tacmb}, \ref{eqn:3}

\noindent
$T_{a,hot}$ = warm dust antenna temperature 

\noindent
$T_{a,ex}$ = extragalactic sources antenna temperature; eqn. \ref{eqn:Tcmb}

\noindent
$T_{a,for}$ = foreground antenna temperature; eqns. \ref{eqn:1-2:Tcmb}, \ref{eqn:Tpallone}

\noindent
$T_{a,gal}$ = Galactic antenna temperature; eqns. \ref{eqn:Tcmb}, \ref{eqn:2-9:Tacmb}

\noindent
$T_{a,gr}$ = ground antenna temperature; eqns. \ref{sky}, \ref{eqn:1-2:Tcmb}, \ref{eqn:2-9:Tacmb}, \ref{eqn:Tpallone}

\noindent
$T_{a,load}$ = antenna temperature of the calibrator; eqns. \ref{eqn:1-2:Ta}, \ref{eqn:2-9:Tacmb}, \ref{tsys}

\noindent
$T_{a,offset}$ = offset antenna temperature; eqn. \ref{eqn:Tpallone}

\noindent
$T_{a,pl}$ = antenna temperature produced by plasma

\noindent
$T_{a,sky}$  = sky antenna temperature; eqn. \ref{sky} 

\noindent
$T_{a,sun}$ = sun antenna temperature; eqns. \ref{sky}, \ref{eqn:1-2:Tcmb}

\noindent
$T_{a,syn}$ = synchrotron emission antenna temperatura; eqn. \ref{eqn:syn}

\noindent
$T_{a,zenith}$ = zenith antenna temperature; eqn. \ref{eqn:1-2:Ta}

\noindent
$T_{a,CMB}$ = CMB antenna temperature; eqns. \ref{eqn:Tcmb}, \ref{eqn:1-2:Tcmb}, \ref{eqn:2-9:Tacmb}, \ref{eqn:Tpallone}

\noindent
$T_{a,H_2O}$ = water vapor antenna temperature; eqn. \ref{eqn:3}

\noindent
$T_{a,HRN}$ = horn antenna temperature; eqn. \ref{eqn:Tpallone}

\noindent
$T_{a,IGS}$ = screen antenna temperature; eqn. \ref{eqn:Tpallone}

\noindent
$T_{a,WND}$ = window antenna temperature; eqn. \ref{eqn:Tpallone}

\noindent
$T_{a,O_2}$ = $O_2$ emission antenna temperature; eqn. \ref{eqn:3}

\noindent
$T_{a,RFI}$ = manmade interfence antenna temperature; eqns. \ref{eqn:1-2:Tcmb}, \ref{eqn:2-9:Tacmb}

\noindent
$T_B$ = broadcast noise temperature

\noindent
$T_{eff}$ = effective temperature of the blackbody calibrator; eqns. \ref{eqn:Ta}, \ref{eqn:teff}

\noindent
$T_{\infty}$ = antenna temperature of the background signal; eqn. \ref{eqn:Tobs}

\noindent
$T_l$ = bolling temperature of the liquid in the dewar; eqn. \ref{eqn:teff}

\noindent
$T_{phys}$ = physical temperature of the components; eqns. \ref{tsys}, \ref{eqn:Tobs}

\noindent
$T_{pl}$ = physical temperature of the plasma

\noindent
$T_r$ = noise temperature radiated by the receiver; eqns. \ref{eqn:Ta}, \ref{eqn:teff}

\noindent
$T_{sys}$ = system temperature; eqn. \ref{tsys}

\noindent
$T^{th}_{CMB}$ =  CMB thermodinamic temperature 

\noindent
$u$ = atmospheric humidity 

\noindent
$y$ = power absorption coefficient of the cable; eqn. \ref{eqn:teff}

\noindent
$\alpha$  = linear absorption coefficient or attenuation 

\noindent
$\alpha_{syn}$ = synchrotron spectral index; eqn. \ref{eqn:syn}.

\noindent
$\gamma_0$ = widht paramenter for the $O_2$ continuum; eqns. \ref{eqn:1}, \ref{eqn:2}

\noindent
$\delta G$ = gain change; eqn. \ref{tsys}

\noindent
$\delta L$ = change in the insertion loss; eqn. \ref{tsys}

\noindent 
$\delta R$ = change in the reflection coefficient of the horn and amplifier; eqn. \ref{tsys}

\noindent
$\delta T_R$ = change in the physical temperature of any loss front-end components

\noindent
$\delta T_{sys}$ = change in the radiometer performance;  eqns. \ref{eqn:2-9:Tacmb}, \ref{tsys}

\noindent
$\Delta T_{a,inst}$ = position-dependent change in receiver output; eqn. \ref{eqn:1-2:Ta}

\noindent
$\Delta T_{a,joint}$ = differential temperature contribution from the imperfect joint between the antenna and the cold load; eqn. \ref{eqn:1-2:Ta}

\noindent
$\epsilon_c$ = coaxial components coefficient 

\noindent
$\epsilon_f$ = flaring section coefficient

\noindent
$\epsilon_w$ = antenna waveguide section coefficient 

\noindent
$\tau_{\nu}$ = optical depth coefficient at frequency $\nu$; eqn. \ref{eqn:Tobs}

\newpage
\null

\newpage
\null

\newpage
\null

\newpage
\null

\newpage
\null

\end{document}